\documentclass[10pt,journal,cspaper,compsoc]{IEEEtran}

\usepackage{graphicx}
\usepackage{graphics,epsfig,subfigure}
\usepackage{color}
\usepackage{amsmath, amsfonts, amssymb
}
\usepackage{algorithm}
\usepackage{algorithmic}
\usepackage{url}
\usepackage{latexsym}
\usepackage{tabularx}
\usepackage{comment}
\usepackage{colortbl}

\newtheorem{theorem}{\bf Theorem}
\newtheorem{lemma}{\bf Lemma}
\newtheorem{cor}{\bf Corollary}

\newcommand{\revision}[1]{{#1}}
\newcommand{\reminder}[1]{{}}

\newcommand{\yy} {\mathbb{Y}}
\newcommand{\zz} {\mathbb{Z}}
\newcommand{\G} {\mathbb{G}}

\newcommand{\dis}       {\mathcal{D}}
\newcommand{\db}{X}
\newcommand{\noisyDB}{Y}
\newcommand{\noisyDBS}{Y_1}
\newcommand{\noisyDBM}{Y_2}

\newcommand{\Noise}{Z}
\newcommand{\NoiseS}{Z_1}
\newcommand{\NoiseM}{Z_2}

\newcommand{\varianceS}{\sigma_1^2}
\newcommand{\varianceM}{\sigma_2^2}

\newcommand{\covMatrix}{K}

\newcommand{\varianceGS}{\sigma_1^2}
\newcommand{\varianceGM}{\sigma_2^2}

\newcommand{\levels}{M}

\newcommand{\myvec} {\mbox{vec}}
\newcommand{\diag} {\mbox{diag}}

\begin{document}

\title{Enabling Multi-level Trust in Privacy Preserving Data Mining}

\author{Yaping~Li,~\thanks{Y. Li was with Department of Electrical
Engineering and Computer Science at University of California at Berkeley, when the work was done; she is now with Department of Information Engineering at the Chinese University of Hong Kong. E-mail: yaping@eecs.berkeley.edu.}~%
Minghua~Chen,~
Qiwei~Li~
and~Wei~Zhang
\thanks{M. Chen, Qiwei Li and Wei Zhang are with Department of Information Engineering at the Chinese University of Hong Kong,
Shatin, Hong Kong, China. E-mail: \{minghua, lqw008, zw007\}@ie.cuhk.edu.hk.} %
}

\markboth{Manuscript accepted for publication in IEEE Transactions on Knowledge and Data Engineering, 2011}%
{Li \MakeLowercase{\textit{et al.}}: Enabling Multi-level Trust in Privacy Preserving Data Mining}

\IEEEcompsoctitleabstractindextext{%
\begin{abstract}
Privacy Preserving Data Mining (PPDM) addresses the problem of
developing accurate models about aggregated data without access to
precise information in individual data record. A widely studied
\emph{perturbation-based PPDM} approach introduces random
perturbation to individual values to preserve privacy before data is
published. Previous solutions of this approach are limited in their
tacit assumption of single-level trust on data miners.

In this work, we relax this assumption and expand the scope of
perturbation-based PPDM to Multi-Level Trust (MLT-PPDM). In our setting, the more trusted a data miner is, the less
perturbed copy of the data it can access. Under this setting, a malicious data miner
may have access to differently perturbed copies of the same
data through various means, and may combine these diverse copies to jointly infer additional information about the original data that the data owner does not intend to release.
Preventing such \emph{diversity attacks} is the key challenge of providing MLT-PPDM services. We address this
challenge by properly correlating perturbation across copies at different
trust levels. We prove that our solution is robust against diversity attacks with respect to our privacy goal. That is, for data miners who have access to an arbitrary collection of the perturbed copies, our solution prevent them from jointly reconstructing
the original data more accurately than the
best effort using any individual copy in the collection. Our solution allows a data owner to generate
perturbed copies of its data for arbitrary trust levels
on-demand. This feature offers data owners maximum flexibility.
\end{abstract}

}
\maketitle
\IEEEpeerreviewmaketitle

\section{Introduction}\label{sec:intro}
Data perturbation, a widely employed and accepted Privacy Preserving Data Mining (PPDM)
approach, tacitly assumes single-level trust on data miners. This approach introduces
uncertainty about individual values before data is published or
released to third parties for data mining
purposes~\cite{aa01:privacy, as00:privacy, cl05:privacy,
hdc05:deriving, lspms07:hiding, lrk06:random, plky07:time}. Under
the single trust level assumption, a data owner generates only one
perturbed copy of its data with a fixed amount of uncertainty. This
assumption is limited in various applications where a data owner
trusts the data miners at different levels.

We present below a two
trust level scenario as a motivating example.
\begin{itemize}
\item The government or a business might do internal (most trusted) data mining, but they
may also want to release the data to the public, and might perturb
it more. The mining department which receives the less perturbed
internal copy also has access to the more perturbed public copy. It would be
desirable that this department does not have {\em more} power in reconstructing the original data
by utilizing both copies than when it has only the internal copy.

\item Conversely, if the internal copy is leaked to the public, then
obviously the public has all the power of the mining department.
However, it would be desirable if the public cannot reconstruct the original data \emph{more} accurately
when it uses both copies than when it uses only the leaked internal copy.
\end{itemize}

This new dimension of {\em Multi-Level Trust} (MLT) poses new
challenges for perturbation based PPDM. In contrast to the
single-level trust scenario where only one perturbed copy is released, now multiple differently perturbed copies
of the same data is available to data miners at different trusted levels. The more trusted a data miner is,
the less perturbed copy it can access; it may also have access to the perturbed copies available at
lower trust levels. Moreover, a data
miner could access multiple perturbed copies through various other means, e.g., accidental leakage or colluding with others.

By utilizing {\emph{diversity} across}
differently perturbed copies, the data miner may be able to
produce a more accurate reconstruction of the original data than what is allowed by the
data owner. We refer to this attack as a~\emph{diversity attack}.
It includes the colluding attack scenario where adversaries combine their copies to
mount an attack; it also includes the scenario where an adversary
utilizes public information to perform the attack on its own.
Preventing diversity attacks is the key challenge in solving
the MLT-PPDM problem.

In this paper, we address this challenge in enabling
MLT-PPDM services. In particular, we focus on the additive
perturbation approach where random Gaussian noise is added to the original
data with \emph{arbitrary} distribution, and provide a systematic solution. Through a one-to-one
mapping, our solution allows a data owner to generate distinctly
perturbed copies of its data according to different trust levels.
Defining trust levels and determining such mappings are beyond the
scope of this paper.

\subsection{Contributions}
We make the following contributions:
\begin{itemize}
    \item We expand the scope of perturbation based PPDM to
    multi-level trust, by relaxing the implicit assumption of
single-level trust in existing work. MLT-PPDM introduces another
dimension of flexibility which allows data owners to generate
differently perturbed copies of its data for different trust levels.

    \item We identify a key challenge in enabling MLT-PPDM
    services. In MLT-PPDM, data miners may have access to multiple perturbed copies.
    By combining multiple
    perturbed copies, data miners may be able to perform diversity attacks to
    reconstruct the original data more accurately than what is
    allowed by the data owner. \revision{Defending such attacks
    is challenging, which we explain through a case study
    in Section~\ref{sec:case.study}. }


    \item We address
    this challenge by properly correlating perturbation across copies at different
    trust levels. We prove that our
    solution is robust against diversity attacks. We propose several algorithms
    for different targeting scenarios. We demonstrate the effectiveness of our solution through
    experiments on real data.

    \item Our solution allows data owners to
generate perturbed copies of their data at arbitrary trust levels
on-demand. This property offers data owners maximum flexibility.
\end{itemize}

\subsection{Related Work}
Privacy Preserving Data Mining (PPDM) was first proposed
in~\cite{as00:privacy} and~\cite{lp00:privacy} simultaneously. To
address this problem, researchers have since proposed various
solutions that fall into two broad categories based on the level of
privacy protection they provide. The first category of the Secure
Multiparty Computation (SMC) approach provides the strongest level
of privacy; it enables mutually distrustful entities to mine their
collective data without revealing anything except for what can be
inferred from an entity's own input and the output of the mining
operation alone~\cite{lp00:privacy,vaidya02:privacy}. In principle,
any data mining algorithm can be implemented by using generic
algorithms of SMC~\cite{goldreich02}. However, these algorithms are
extraordinarily expensive in practice, and impractical for real use.
To avoid the high computational cost, various solutions that are
more efficient than generic SMC algorithms have been proposed for
specific mining tasks. Solutions to build decision trees over the
horizontally partitioned data were proposed in~\cite{lp00:privacy} .
For vertically partitioned data, algorithms have been proposed to
address the association rule mining~\cite{vaidya02:privacy},
$k$-means clustering~\cite{vc03:privacypreserving}, and frequent
pattern mining problems~\cite{fww05:privacy}. The work
of~\cite{bagzcca06:using} uses a secure coprocessor for privacy
preserving collaborative data mining and analysis.

The second category of the partial information hiding approach trades
privacy with improved performance in the sense that malicious data
miners may infer certain properties of the original data from the
disguised data. Various solutions in this category allow a data
owner to transform its data in different ways to hide the true
values of the original data while at the same time still permit
useful mining operations over the modified data. This approach can
be further divided into three categories: (a)
$k$-anonymity~\cite{ay04:condensation,bcyd05:privacy,
kg06:privacy,mgkv06:ldiversity, s02:model,xt06:privacy}, (b)
retention replacement (which retains an element with probability $p$
or replaces it with an element selected from a probability
distribution function on the domain of the
elements)~\cite{ast05:privacy,dz03:using,esag02:assoc}, and (c) data
perturbation (which introduces uncertainty about individual values
before data is published)~\cite{aa01:privacy, as00:privacy,
cl05:privacy, hdc05:deriving, lspms07:hiding,
lrk06:random, plky07:time, kdws03:privacy}.


The data perturbation approach includes two main classes of methods:
additive~\cite{aa01:privacy, as00:privacy, hdc05:deriving,
lspms07:hiding, plky07:time} and matrix multiplicative~\cite{cl05:privacy,
lrk06:random} schemes. These methods apply mainly to continuous
data. In this paper, we focus solely on the additive perturbation
approach where noise is added to data values.




Another relevant line of research concerns the problem of privately
computing various set related operations. Two party protocols for
intersection, intersection size, equijoin, and equijoin size were
introduced in~\cite{aes03:ii} for honest-but-curious adversarial
model. Some of the proposed protocols leak
information~\cite{rakesh06:sovereign}. Similar protocols for set
intersection have been proposed in~\cite{clifton03tools,
hfh99:enhancing}. Efficient two party protocols for the private
matching problem which are both secure in the malicious and
honest-but-curious models were introduced
in~\cite{freedman04efficient}. Efficient private and threshold set
intersection protocols were proposed
in~\cite{kissner-privacypreserving}. While most of these protocols
are equality based, algorithms in ~\cite{rakesh06:sovereign} compute
arbitrary join predicates leveraging the power of a secure
coprocessor. Tiny trusted devices were used for secure function
evaluation in~\cite{A05:more}.

Our work does not re-anonymizing a dataset after it is updated with insertions and/or deletions, which is a topic studied by the authors in \cite{byun2006secure, xt07:minvariance, fung2008:acd, wang2008inference}. Instead, we study anonymizing the same dataset at multiple trust levels. The two problems are orthogonal.

An earlier version of this paper appeared in~\cite{li:multilevel.trust}
and initiated the topic of MLT-PPDM. Recently, Xiao
et al. proposed an algorithm of multi-level uniform
perturbation~\cite{xiao:perturbation.multiple}. Our paper differs from~\cite{xiao:perturbation.multiple} in three
main aspects. Firstly, the two papers address different problems and tackle the problems under different privacy measures. We propose multi-level privacy preserving for additive Gaussian noise perturbation, and use a measure based on how closely the original values can be reconstructed from the perturbed data~\cite{as00:privacy,hdc05:deriving,lspms07:hiding}. While~\cite{xiao:perturbation.multiple} presents an algorithm of multi-level uniform perturbation, and studies its performance using the $\rho_1-\rho_2$ privacy measure~\cite{egs03limiting}. As a result, neither the solution in~\cite{xiao:perturbation.multiple} can be easily
applied to the problem in this paper nor the solution in this paper can be directly applied to the problem in~\cite{xiao:perturbation.multiple}.
Secondly, based on Gaussian noise perturbation, the solution in this paper is more suitable for high-dimensional data,
as compared to that in~\cite{xiao:perturbation.multiple} based on uniform perturbation~\cite{aggarwal2008privacy}.
Thirdly, We present several nontrivial theoretical
results. We discuss reconstruction errors under independence
noise, analyze the security of our scheme when
collusion occurs, and study the computational complexities
based on Kroneckor product. These results provide fundamental insights into the problem.

\subsection{Paper Layout}
The rest of the paper is organized as follows. We go over preliminaries in Section~\ref{sec:preliminaries}.
We formulate the problem, and define our privacy goal in Section~\ref{sec:prob.form}.
In Section~\ref{sec:case.study}, we present a simple but important case study.
It highlights the key challenge in achieving our privacy goal, and
presents the intuition that leads to our solution. In Section~\ref{sec:solution}, we formally present
our solution, and prove that it achieves our privacy goal. Algorithms that target different scenarios are also proposed,
and their complexities are studied.
We carry out extensive experiments on real data in Section~\ref{sec:simulation}
to verify our theoretical analysis. Section~\ref{sec:conclusion} concludes the paper.

\section{Preliminaries}\label{sec:preliminaries}

\subsection{Jointly Gaussian}\label{ssec:jnt.gaussian}
In this paper, we focus on perturbing data by additive
Gaussian noise~\cite{aa01:privacy, as00:privacy, hdc05:deriving,
lspms07:hiding, plky07:time}, i.e., the added noises are jointly Gaussian.\footnote{Note that we do not make
any assumptions about the distribution of the data.}

Let $G_1$ through $G_L$ be $L$ Gaussian random variables.
They are said to be \emph{jointly Gaussian} if and only if each of
them is a linear combination of multiple independent Gaussian random
variables.\footnote{Two random variables are independent if knowing
the value of one yields no knowledge about that of
the other. Mathematically, two random variables $G_1$ and $G_2$ are
independent if, for any values $g_1$ and $g_2$, $f_{G_1,G_2} (g_1, g_2) = f_{G_1} (g_1) f_{G_2} (g_2)$, where $f_{G_1,G_2} (g_1, g_2)$ is the joint probability density function
of $G_1$ and $G_2$, and $f_{G_1}(g_1)$ and $f_{G_2}(g_2)$ are the probability density functions of $G_1$ and $G_2$, respectively.
Generally, random variables $G_1$ through $G_L$ are mutually independent if, for any values $g_1$ through $g_L$, $f_{G_1, ..., G_L} (g_1, ..., g_L) = f_{G_1} (g_1) ... f_{G_L} (g_L)$.} Equivalently, $G_1$ through $G_L$ are
jointly Gaussian if and only if any linear combination of them is
also a Gaussian random variable.

A vector formed by jointly Gaussian random variables is called a
jointly Gaussian vector. For a jointly Gaussian vector $\G=[G_1,
\ldots, G_L]^T$, its probability density function (PDF) is as
follows: for any real vector $g$,
\[
    f_{\G}(g) = \frac{1}{\sqrt{(2\pi)^L \det (K_{\G})}}e^{-(g-\mu_{\G})^TK^{-1}_{\G}(g-\mu_{\G})/2},
\]
where $\mu_{\G}$ and $K_{\G}$ are the mean vector and covariance matrix of
${\G}$, respectively.

Note that not all Gaussian random variables are jointly Gaussian.
For example, let $G_1$ be a zero mean Gaussian random variable with
a positive variance, and define $G_2$ as
\[
    G_2 = \left\{
            \begin{array}{ll}
              G_1, & \hbox{if $|G_1|\leq 1$;} \\
              -G_1, & \hbox{otherwise.}
            \end{array}
          \right.
\]
where $|G_1|$ is the absolute value of $G_1$. It is straightforward to verify that $G_2$ is Gaussian, but
$G_1+G_2$ is not. Therefore, $G_1$ and $G_2$ are not jointly
Gaussian.

If multiple random variables are jointly Gaussian, then conditional on a subset of them, the remaining variables are still jointly Gaussian.
Specifically, partition a jointly Gaussian vector $\G$ as
$$\G = \begin{bmatrix}
\G_1 \\
\G_2 \\
\end{bmatrix}$$ and $$\mu_{\G} = \begin{bmatrix}
\mu_1 \\
\mu_2 \\
\end{bmatrix}, ~~ K_{\G} = \begin{bmatrix}
K_{11} & K_{12} \\
K_{21} & K_{22} \\
\end{bmatrix}$$accordingly. Then the distribution of $\G_2$ given $\G_1 = v_1$ is also a jointly Gaussian
with mean $\mu_2 + K_{21} K_{11}^{-1} (v_1 - \mu_1)$ and covariance matrix $K_{22} - K_{21} K_{11}^{-1} K_{21}^T$ \cite[ch 2.5]{shanmugan1988random}.
This is a key property of jointly Gaussian variables. We utilize this property in Section~\ref{ssec:on.the.fly}.

\subsection{Additive Perturbation}\label{ssec:add.prtub}
The single-level trust PPDM problem via data perturbation has been widely
studied in literature. In this setting, a data owner implicitly
trusts all recipients of its data uniformly and distributes a single
perturbed copy of the data.

A widely used and accepted way to perturb data is by additive
perturbation~\cite{aa01:privacy, as00:privacy, hdc05:deriving,
lspms07:hiding, plky07:time}. This approach adds to the original
data, $\db$, some random noise, $\Noise$, to obtain the perturbed
copy, $\noisyDB$, as follows:
\begin{equation}\label{eqn:sngl.lvl}
        \noisyDB = \db + \Noise.
\end{equation}
We assume that $\db$, $\noisyDB$, and $\Noise$ are all $N$-dimension
vectors where $N$ is the number of attributes in $\db$. Let $x_j,
y_{j}$, and $z_{j}$ be the $j^{th}$ entry of $\db$, $\noisyDB$, and
$\Noise$ respectively.

The original data $\db$ follows a distribution with mean vector
$\mu_X$ and covariance matrix $K_X$. The covariance $K_{X}$ is an $N
\times N$ positive semi-definite matrix given by
\begin{equation}
        K_{X} = E\left[(X-\mu_X) (X-\mu_X)^T\right],
\end{equation}
which is a diagonal matrix if the attributes in $\db$ are
uncorrelated.

The noise $\Noise$ is assumed to be independent of $\db$ and is a
jointly Gaussian vector with zero mean and covariance matrix $K_Z$
chosen by the data owner. In short, we write it as $\Noise \sim
N(0,K_Z)$. The covariance matrix $K_{Z}$ is an $N \times N$ positive
semi-definite matrix given by
\begin{equation}
        K_{Z} = E\left[Z Z^T\right].
\end{equation}

It is straightforward to verify the mean vector of $Y$ is also
$\mu_X$, and its covariance matrix, denoted by $K_Y$, is
\[
    K_Y = K_X + K_Z.
\]

The perturbed copy $\noisyDB$ is published or released to data
miners. Equation~\ref{eqn:sngl.lvl} models both the cases where a
data miner sees a perturbed copy of $\db$, and where it knows the
true values of certain attributes. The latter scenario is considered
in recent work~\cite{plky07:time} where the authors show that
sophisticated filtering techniques utilizing the true value leaks
can help recover $\db$.


In general, given $\noisyDB$, a malicious data miner's goal is to
reconstruct $\db$ by filtering out the added noise. The authors
of~\cite{hdc05:deriving} point out that the attributes in
$\db$ and the added noise should have the same correlation,
otherwise the noise can be easily filtered out.
This observation essentially requires to choose $K_Z$ to be
proportional to $K_X$~\cite{hdc05:deriving}, i.e., $K_Z = \sigma_Z^2 K_X$ for some constant $\sigma_Z^2$ denoting the perturbation magnitude.

\subsection{Linear Least Squares Error Estimation}\label{ssec:LLSE}
Given a perturbed copy of the data, a malicious data miner may attempt to
reconstruct the original data as accurately as possible. Among
the family of linear reconstruction methods, where estimates can
only be linear functions of the perturbed copy, \emph{Linear Least
Squares Error} (LLSE) estimation has the minimum square
errors between the estimated values and the original values \cite[ch 7.1--7.2]{shanmugan1988random}.



The LLSE estimate of $\db$ given $\noisyDB$, denoted by
$\hat{\db}(\noisyDB)$, is (see Appendix \ref{appendix:proof.eqn.llse1} for the deduction)
\begin{eqnarray}
    \hat{\db}(\noisyDB) = \covMatrix_{\db\noisyDB}
    \covMatrix^{-1}_{\noisyDB} \left(\noisyDB-\mu_X\right) +\mu_X, \label{eqn:llse1}
\end{eqnarray}
where $\covMatrix_{\db\noisyDB}$ ($\covMatrix_{\noisyDB}$ resp.) is
the covariance matrix of $\db$ and $\noisyDB$ ($\noisyDB$ resp.).
$\covMatrix_{\db\noisyDB}$ is given by
\begin{eqnarray}
    \covMatrix_{\db\noisyDB}&=&E[(\db-\mu_X)(\noisyDB-E[\noisyDB])^T]\nonumber\\
    &=&E[(\db-\mu_X)((\db-\mu_X)+(\Noise-0))^T]\nonumber\\
    &=&\covMatrix_{\db}+0 = \covMatrix_{\db}.\nonumber
\end{eqnarray}
\revision{Note in the above derivation, we compute $E[(\db-\mu_X)Z^T]=E[(\db-\mu_X)]E[Z^T]=0$, since $\db$ and $Z$ are independent.}

The square estimation errors between the LLSE estimates and the
original values of the attributes in $\db$ are the diagonal
terms of the covariance matrix of $\db- \hat{\db}(\noisyDB)$. An important property of LLSE estimation is that
it simultaneously minimizes all these estimation errors.

\subsection{Kronecker Product}
In the MLT-PPDM problem, the covariance matrix of noises can be written as the Kronecker product \cite{brewer1978kronecker} of two matrices. In this paper, we explore the properties of the Kronecker product for efficient computation.

The Kronecker product \cite{brewer1978kronecker} is a binary matrix operator that maps two matrices of arbitrary dimensions into
a larger matrix with a special block structure. Given an $n \times m$ matrix $A$ and $p \times q$ matrix $B$,
where $$A = \begin{bmatrix}
a_{11} & \cdots & a_{1m} \\
\vdots & \ddots & \vdots \\
a_{n1} & \cdots & a_{nm}
\end{bmatrix},$$
their Kronecker product, denoted as $A \otimes B$, is an $np \times mq$ matrix with the block structure
$$\begin{bmatrix}
a_{11} B & \cdots & a_{1m} B \\
\vdots & \ddots & \vdots \\
a_{n1} B & \cdots & a_{nm} B
\end{bmatrix}.$$

We list several properties of Kronecker product that will be used later. Assume that $A$, $B$, $C$ and $D$ are matrices
and their dimensions are appropriate for the computation in each property, we have

\begin{enumerate}
\item $(\alpha A) \otimes B = A \otimes (\alpha B) = \alpha (A \otimes B)$, where $\alpha \in \mathbb{R}$;
\item $(A \otimes B)^T = A^T \otimes B^T$;
\item $(A \otimes B)^{-1} = A^{-1} \otimes B^{-1}$;
\item $(A \otimes B)(C \otimes D) = AC \otimes BD$;
\item $\myvec(A B C) = (C^T \otimes A) \myvec(B)$, where $\myvec(\cdot)$ denotes the vectorization of a matrix formed by stacking the columns of the matrix into a single column vector.
\end{enumerate}

\section{Problem Formulation}\label{sec:prob.form}
In this section, we present the problem settings,
describe our threat model, state our
privacy goal, and identify the design space. Table~\ref{tab:notation} lists the
key notations used in the paper.
\begin{table}[t]
\caption{Key Notations}%
\label{tab:notation}
\begin{center}%
\begin{tabular}
[c]{l|l}\hline \textbf{Notation} & \textbf{Definition}\\\hline
$\db$ & original data\\
$Y_i$ & perturbed copy of $\db$ of trust level $i$\\
$Z_i$ & noise added to $\db$ to generate $Y_i$\\
$\levels$ & number of trust levels\\
$N$ & number of attributes in $\db$\\
$\yy$ & a vector of all $M$ perturbed copies \\
$\zz$ & a vector of noise $\Noise_1$ to $\Noise_{\levels}$ \\
$\hat{X}(\yy)$ & LLSE estimate of $X$ given $\yy$ \\
$K_X$ & covariance matrix of $\db$ \\
$K_{\zz}$ & covariance matrix of $\zz$\\\hline
\end{tabular}
\end{center}
\end{table}

\subsection{Problem Settings}\label{ssec:prob.setting}
\revision{In the MLT-PPDM problem we consider in this paper, a
data owner trusts data miners at different levels and generates a series of perturbed copies of its data
for different trust levels. This is done by adding
varying amount of noise to the data.

Under the multi-level trust setting, data miners at higher trust levels can access less perturbed copies. Such less perturbed copies are not accessible by data miners at lower trust levels. In some scenarios, such as the motivating example we give at the beginning of Section~\ref{sec:intro}, data miners at higher trust levels may also have access to the perturbed copies at more than one trust levels. Data miners at different trust levels may also collude to share the perturbed copies among them. As such, it is common that data miners can have access to more than one perturbed copies.}

Specifically, we assume that the data owner wants to release
$\levels$ perturbed copies of its data $\db$, which is an $N\times
1$ vector with mean $\mu_{\db}$ and covariance $K_{\db}$ as defined
in Section~\ref{ssec:add.prtub}. These $\levels$ copies can be
generated in various fashions. They can be jointly generated all at
once. Alternatively, they can be generated at different times upon
receiving new requests from data miners, in an on-demand fashion.
The latter case gives data owners maximum flexibility.

It is true that the data owner may consider to release only the mean and covariance of the original data.
We remark that simply releasing the mean and covariance does not provide the same utility as the perturbed data.
For many real applications, knowing only the mean and covariance may not be sufficient to apply data mining techniques,
 such as clustering, principal component analysis, and classification \cite{lrk06:random}.
By using random perturbation to release the dataset, the data owner allows the data miner to exploit more statistical information without releasing
the exact values of sensitive attributes \cite{aa01:privacy, as00:privacy}.


Let $\yy=[Y_1^T,\ldots, Y^T_M]^T$ be the vector
of all perturbed copies $Y_i (1\leq i\leq M)$. Let $\zz = [\Noise^T_1, \ldots, \Noise^T_M]^T$ be the
vector of noise. Let $H$ be an $(N\cdot M)\times N$
matrix as follows:
\[
    H = \left[\begin{array}{c}
        I_{N}\\
        \vdots\\
        I_{N}\end{array}\right],
\]
where $I_N$ represents an $N\times N$ identity matrix.

We have the relationship between $\yy$, $\db$ and $\zz$ as follows:
\begin{equation}
            \yy = \left[\begin{array}{c}
    Y_{1}\\
    \vdots\\
    Y_{M}\end{array}\right]=\left[\begin{array}{c}
        I_{N}\\
        \vdots\\
        I_{N}\end{array}\right]X+\left[\begin{array}{c}
    Z_{1}\\
    \vdots\\
    Z_{M}\end{array}\right] = HX +\zz,
\end{equation}
where $\Noise_i, 1\leq i \leq M$ are independent of $\db$. To be
robust against advanced filtering attacks, individual
noise terms in $Z_i$ added to different attributes in
$\db$ should have the same correlations as the attributes
themselves, otherwise $Z_i$ can be easily filtered
out~\cite{hdc05:deriving}. As such, we have
\[
    K_{Z_i} = \sigma_{Z_i}^2 K_X, \mbox{ and } K_{Y_i} = (1+\sigma_{Z_i}^2)K_X,
\]
where $\sigma_{Z_i}^2$ is a constant of the perturbation
magnitude. The data owner chooses a value for $\sigma_{Z_i}^2$
according to the trust level associated with the target perturbed
copy $Y_i$.



\subsection{Threat Model}\label{ssec:threat-model}
We assume malicious data miners who always attempt to
reconstruct a more accurate estimate of the original data given
perturbed copies. We hence use the terms data miners and
adversaries interchangeably throughout this paper.
In MLT-PPDM, adversaries may have access to a subset of the perturbed copies of the data. The adversaries' goal is to reconstruct the original data as accurately as possible based on all available perturbed copies.

The reconstruction accuracy depends heavily on the adversaries'
knowledge. We make the same assumption as the one in~\cite{hdc05:deriving} that adversaries have the knowledge
of the statistics of the original data $X$ and the noise
$\zz$, i.e., mean $\mu_X$, and covariance matrices $K_X$ and
$K_{\zz}$. Note the adversaries with less knowledge are weaker than the ones we study in this paper.


\revision{
In addition, we assume adversaries only perform linear estimation
attacks, where estimates can only be linear functions of the
perturbed data $\noisyDB$. It is known that if $\db$ follows a jointly Gaussian distribution, then
LLSE estimation achieves the minimum estimation error among both
linear and nonlinear estimation methods. For $\db$ with general distribution, LLSE
estimation has the minimum estimation error among all linear
estimation methods. Various recent work in perturbation based
PPDM, such as~\cite{hdc05:deriving} and~\cite{lspms07:hiding}, makes this assumption of
linear estimation. See reference~\cite{plky07:time} for a comprehensive review.

Noticed $K_{\db \yy}=K_XH^T$ and $K_{\yy}=HK_XH^T + K_{\zz}$, the LLSE estimate $\hat{\db}(\yy)$ of $\db$ given $\yy$ can be expressed as:
\begin{eqnarray}\label{eqn:LLSE.yy}
    \hat{\db}(\yy)
    &=& K_{\db \yy}K^{-1}_{\yy}\left(\yy - E[\yy]\right) +\mu_X \nonumber \\
    &=& K_{X}H^T\left[HK_XH^T + K_{\zz}\right]^{-1}\left(\yy - H\mu_X\right) \nonumber \\
    && +\mu_X.\;\;\;\;\;
\end{eqnarray}}
In our setting, $\hat{\db}(\yy)$ is the most accurate estimate of
$\db$ that an adversary can possibly make. The corresponding
estimation errors of attributes in $\db$ are the diagonal terms of
the covariance matrix of $\hat{\db}(\yy) - \db$. Using Equation~\ref{eqn:LLSE.yy}, we can compute the covariance matrix as follows:
\begin{eqnarray}\label{eqn:LLSE.yy.est.err}
    && E\left[\left(\hat{\db}(\yy) - \db\right)\left(\hat{\db}(\yy) - \db\right)^T\right] \nonumber \\
    &=& K_X - K_X H^T K^{-1}_{\yy}H K_X = \left[K_X^{-1} + H^T K^{-1}_{\zz} H\right]^{-1}. \;\;\;\;\;
\end{eqnarray}
\revision{For an adversary who observes only a single copy $Y_i\, (1\leq i\leq M)$ and gets a LLSE estimate $\hat{\db}(Y_i)$, the covariance matrix of $\hat{\db}(Y_i) - \db$ has a simple form as follows:
\begin{eqnarray}\label{eqn:LLSE.Y.est.err}
    && E\left[\left(\hat{\db}(Y_i) - \db\right)\left(\hat{\db}(Y_i) - \db\right)^T\right] \nonumber\\
    &=& K_X - K_X K^{-1}_{Y_i}K_X = \frac{\sigma^2_{Z_i}}{\sigma^2_{Z_i}+1}K_X.
\end{eqnarray}
}


\subsection{Definitions}\label{ssec:def}

\subsubsection{Distortion}\label{sssec:dis}
To facilitate future discussion on privacy, we define
the concept of perturbation $\dis$ between two datasets as the
average expected square difference between them. For example, the
distortion between the original data $\db$ and the perturbed
copy $\noisyDB$ as defined in Section~\ref{ssec:add.prtub} is
given by:
$$
        \dis(\db, \noisyDB) = \frac{1}{N}\sum_{j=1}^N E[(y_j -
        x_j)^2]\geq 0.
$$
It is easy to see that $\dis(\db, \noisyDB)= \dis(\noisyDB, \db)$.

Based on the above definition, we refer to a perturbed copy
$\noisyDBM$ to be {\em more perturbed} than $\noisyDBS$ with respect
to $X$ if and only if $\dis(\db,\noisyDBM) > \dis(\db,\noisyDBS)$.

\subsubsection{Privacy under Single-level Trust Setting} \label{sssec:priv.util.single}

With respect to the original data $X$, the privacy of a
perturbed copy $Y$ represents how well the true values of
$X$ is hidden in $Y$.

A more perturbed copy of the data does not
necessarily have more privacy 
since the added noise may be intelligently filtered out.
Consequently, we define the privacy of a perturbed copy by
taking into account an adversary's power in reconstructing the
original data. We define the {\em privacy} of $\noisyDB$ with
respect to $\db$ to be $\dis(\db, \hat{\db}(Y))$, i.e., the
distortion between $\db$ and the LLSE estimate $\hat{\db}(Y)$. A larger distortion hides the
original values better (and thus preserves more privacy), so we
refer to a perturbed data $\noisyDBM$ to preserve {\em more privacy}
than $\noisyDBS$ with respect to $\db$ if and only if $\dis(\db,
\hat{\db}(\noisyDBM)) > \dis(\db, \hat{\db}(\noisyDBS))$.

\subsubsection{Privacy under Multi-level Trust Setting}\label{sssec:mprivacy}
We now define privacy for the multi-level trust case
in the same spirit of the single-level trust case.

For a vector $\yy =[\noisyDBS^T, \cdots, \noisyDB_{\levels}^T]^T$
of $M$ perturbed copies of $\db$, the privacy
of $\yy$ represents how well the true values of $X$ is
hidden in the multiple perturbed copies $\yy$. 
The privacy of $\yy$, with
respect to $\db$, is defined as $ \dis(\db, \hat{\db}(\yy))$, the
distortion between $\db$ and its LLSE estimate $\hat{\db}(\yy)$. 


\subsection{Privacy Goal and Design Space}\label{ssec:goal}
In a MLT-PPDM setting, a data owner releases distinctly
perturbed copies of its data to multiple data miners. One key goal
of the data owner is to control the amount of information about its
data that adversaries may derive.


We assume that the data owner wants to distribute a total
of $\levels$ different perturbed copies of its data, i.e., $Y_i
(1\leq i\leq M)$, each for a trust level $i$.
The assumption of $\levels$ is for ease of analysis. It will become clear later that our solution of the
on-demand generation allows a data owner to generate as many
different copies as it wishes.


The data owner can easily control the amount of the
information about its data an attacker may infer from a single
perturbed copy. Utilizing Equation~\ref{eqn:LLSE.Y.est.err}, we
express the privacy of $Y_i$, i.e., $\dis(\db,
\hat{\db}(Y_i))$, as follows:
\begin{eqnarray}\label{eqn:sngl.lvl.pvcy}
&& \dis(\db, \hat{\db}(Y_i)) \nonumber\\
    &=& \frac{1}{N}Tr\left(E\left[\left(\hat{\db}(Y_i) - \db\right)\left(\hat{\db}(Y_i) - \db\right)^T\right]\right) \nonumber \\
    &=& \frac{\sigma^2_{Z_i}}{\sigma^2_{Z_i}+1}\frac{1}{N}Tr\left(K_X\right),
\end{eqnarray}
where $Tr(\cdot)$ represents the trace of a matrix.

The data owner can easily control the privacy 
of an individual copy $Y_i$ by setting $\sigma^2_{Z_i}$ according to trust
level $i$ through a one-to-one mapping. Defining trust levels and
such mappings are beyond the scope of this paper.

However, such control alone is not sufficient in the face
of diversity attacks. Adversaries that can access copies at different trust levels enjoy
the diversity gain when they combine multiple distinctly perturbed
copies to estimate the original data. We discuss one such case in
Section~\ref{ssec:case.ind.noise}.



Ideally, the amount of information about $X$ that
 adversaries can jointly infer from multiple perturbed copies should be no more than that
of the best effort using any individual copy.

Formally, we say the privacy goal is achieved with respect to $M$
perturbed copies $Y_i, 1\leq i\leq M$, if the following statement
holds. For an \emph{arbitrary} subset $\yy_C$ of $\{Y_i, 1\leq i\leq M\}$,
\begin{equation}\label{eqn:goal}
    \dis(\db, \hat{\db}(\yy_C)) = \min_{\xi \in \yy_C} \dis(\db, \hat{\db}(\xi)).
\end{equation}
where $\yy_C$ is the set of perturbed copies an adversary uses to reconstruct the original data.

Intuitively, achieving the privacy goal requires that given the copy
with the least privacy 
among any subset of these $M$ perturbed
copies, the remaining copies in that subset contain no extra
information about $X$.

To achieve this goal, the available design space is noise $\zz$. We
already determine that individual noise $Z_i, 1\leq i\leq M$ must
follow $N(0, \sigma^2_{Z_i}K_X)$. In the rest of the
paper, we show by properly correlating
noise $Z_i, 1\leq i\leq M$, the desired privacy goal can be
achieved.

\section{Case Study}\label{sec:case.study}
In this section, we study a basic case corresponding to the motivating example we described at the beginning of Section~\ref{sec:intro}. In the case, a data miner has access to two differently perturbed copies of the same data, each for a different trust level. We present the challenges in achieving the privacy goal in Equation~\ref{eqn:goal} with two false starts. As we develop a solution to this basic base, we show the key ideas in solving the more general case of
arbitrarily fine granularity of trust levels.

\subsection{An Illustrative Case}
For ease of illustration, we assume single attribute data. We
assume that the data owner has already distributed a perturbed copy
$\noisyDBM$ of the original data $\db$ where
\[
    \noisyDBM = \db + \NoiseM.
\]
Denote the variance of $X$ as $\sigma_X^2$, and the Gaussian noise
$\NoiseM \sim N(0, \varianceM \sigma_X^2)$ is independent of
$\db$.

The data owner now wishes to produce another perturbed copy
$\noisyDBS$. It generates Gaussian noise $\NoiseS \sim N(0,
\varianceS \sigma_X^2)$, and adds it to $\db$ to obtain $\noisyDBS$ as
\[
    \noisyDBS = \db + \NoiseS.
\]
The new noise $\NoiseS$ is also independent of $\db$ (but could be
designed to be correlated with $\NoiseM$). We consider the case where the data owner chooses $\varianceM > \varianceS$ so that $\noisyDBS$ is less perturbed than $\noisyDBM$.

The privacy goal in Equation~\ref{eqn:goal} requires that
\begin{equation}\label{eqn:case.goal}
\dis(\db, \hat{\db}(\noisyDBS, \noisyDBM)) =
\dis(\db, \hat{\db}(\noisyDBS)).
\end{equation}
To see this, note that $\min (\dis(\db, \hat{\db}(\noisyDBS)),
\dis(\db, \hat{\db}(\noisyDBM)))$ can be simplified to $\dis(\db,
\hat{\db}(\noisyDBS))$, i.e., the less perturbed copy gives
better estimate.


\subsection{Two False Starts}
In this section, we illustrate the challenges in achieving the
privacy goal with two false starts.

\subsubsection{Independent Noise}\label{ssec:case.ind.noise}
The first intuitive attempt is to generate the two perturbed
copies independently. The added noise in the two perturbed copies
is not only independent to the original data, but also independent
to each other.

In the case we consider, the above solution generates $\NoiseS$ to
be independent of $\db$ and $\NoiseM$ respectively. Consequently,
adversaries have two perturbed copies as follows:
\begin{equation*}
    \left\{
      \begin{array}{l}
        \noisyDBS = \db + \NoiseS \\
        \noisyDBM = \db + \NoiseM
      \end{array}
    \right.
\end{equation*}
where $\db$, $\NoiseS$ and $\NoiseM$ are mutually
independent. The adversaries perform a joint LLSE estimation to
obtain $\hat{\db}(\noisyDBS, \noisyDBM)$. Straightforward
computation utilizing Equation \ref{eqn:LLSE.yy.est.err} shows that
$$
        \dis(\db, \hat{\db}(\noisyDBS, \noisyDBM)) = \frac{\sigma^2_X}
        {1+1/\varianceS+
        1/\varianceM}.
$$
This value is strictly smaller than the error of the estimate based
on either $\noisyDBS$ or $\noisyDBM$, which is for $i=1, 2$,
$$
        \dis(\db, \hat{\db}(Y_i)) = \frac{\sigma^2_X}{1+1/\sigma^2_i
        },
$$following Equation \ref{eqn:LLSE.Y.est.err}.
Thus, Equation~\ref{eqn:case.goal} is not satisfied and the desired
privacy goal is not achieved.

\textbf{Example}. Assume that the original dataset has single attribute data $\db$
with mean $\mu_X = 10$ and variance $\sigma_X^2 = 1$. The data owner releases perturbed copies $\noisyDBS = \db+\NoiseS$ and $\noisyDBM = \db+\NoiseM$
of two (sensitive) values $\db = [9, 11]^T$ to Alice and Bob
with different trust levels $\varianceS = 1$ and $\varianceM = 4$, respectively.

Alice reconstructs the data values using Eqn. (\ref{eqn:llse1}), and obtains
$\hat{\db}(\noisyDBS) = [9.5, 10.5]^T + 0.5\NoiseS$. The average estimation error is
$$\frac{1}{2} E[(\hat{\db} - \db)^T (\hat{\db} - \db)]
= 0.125 E[\NoiseS^T \NoiseS] + 0.25 = 0.5.$$

Bob reconstructs the data values using Eqn. (\ref{eqn:llse1}), and obtains
$\hat{\db}(\noisyDBS) = [9.8, 10.2]^T + 0.2\NoiseM$. The average estimation error is
$$\frac{1}{2} E[(\hat{\db} - \db)^T (\hat{\db} - \db)]
= 0.02 E[\NoiseM^T \NoiseM] + 0.64 = 0.8.$$

Assume that $\noisyDBS$ and $\noisyDBM$ are generated independently.
The reconstructed data after the collusion between Alice and Bob using Eqn. (\ref{eqn:LLSE.yy})
are $\hat{\db}(\yy) = [85,95]^T/9 + 4\NoiseS/9 + \NoiseM/9$. The average estimation error is
$$\frac{1}{2} E[(\hat{\db} - \db)^T (\hat{\db} - \db)]
= \frac{8}{81} \NoiseS^T \NoiseS + \frac{1}{162} \NoiseM^T \NoiseM + \frac{16}{81} = \frac{4}{9}.$$
Thus the collusion results in a smaller error. \hfill $\Box$

Intuitively, this is because the two copies of the data are
generated independently, each containing some innovative information
of the original data that is absent from the other. When estimation
is performed jointly, the innovative information from both copies
can be utilized, resulting in a smaller estimation error and thus a
more accurate estimate.

\subsubsection{Linearly Dependent Noise}
In light of the incorrectness of the first solution, one might
consider a second approach to generate new noise so that it is linearly dependent to the existing one.

In the case we consider, the above approach may generate $\NoiseS =
\frac{\sigma_1}{\sigma_2}\NoiseM$. It is easy to verify that
$\NoiseS\sim N(0, \sigma_1^2 \sigma_X^2)$. However, $\noisyDBS=\db + \NoiseS$
again fails to achieve the privacy goal.

To see this, notice that the adversaries who have access to both copies can reconstruct
$\db$ perfectly as follows:
$$
\db = \frac{\sigma_2 \noisyDBS - \sigma_1 \noisyDBM}{\sigma_2-\sigma_1} =\frac{\sigma_2 (\db+\NoiseS)-\sigma_1(\db+\NoiseM)}{\sigma_2-\sigma_1}.
$$
The estimation error is zero, and Equation~\ref{eqn:case.goal} is not
satisfied.

\subsection{Proposed Solution}
Intuitively, Equation~\ref{eqn:case.goal} requires that given
$\noisyDBS$, observing the more perturbed $\noisyDBM$ does not
improve the estimation accuracy.

One way to satisfy Equation~\ref{eqn:case.goal} is to generate
$\NoiseS$ so that $\noisyDBS= \db+\NoiseS$ and $\NoiseM-\NoiseS$ are
independent. To see why, we rewrite $\noisyDBM$ as
\begin{equation} \label{eqn:z2minusz1}
    \noisyDBM = \noisyDBS + (\NoiseM - \NoiseS).
\end{equation}
If $\noisyDBS$ and $\NoiseM-\NoiseS$ are independent, then
$\noisyDBM$ is nothing but a perturbed observation of $\noisyDBS$.
All information in $\noisyDBM$ useful for estimating $\db$ is
inherited from $\noisyDBS$. Consequently, given $\noisyDBS$,
$\noisyDBM$ provides no extra innovative information to improve the
estimation accuracy, and Equation~\ref{eqn:case.goal} is satisfied.

Since $\db$ and $\NoiseS$ (resp. $\NoiseM$) are independent,
$\noisyDBS$ and $\NoiseM-\NoiseS$ are independent if $\NoiseS$ and
$\NoiseM-\NoiseS$ are independent. The following theorem gives a
sufficient and necessary condition for $\NoiseS$ and
$\NoiseM$ to satisfy that $\NoiseS$ and $\NoiseM-\NoiseS$ are
independent.
\begin{theorem}\label{theorem:case.independence}
Assume $\NoiseS \sim N(0, \sigma^2_1 \sigma_X^2)$, $\NoiseM \sim N(0,
\sigma^2_2 \sigma_X^2)$, and $\sigma_1^2<\sigma^2_2$. $\NoiseS$ and
$\NoiseM-\NoiseS$ are independent if and only if $\NoiseS$
and $\NoiseM$ are jointly Gaussian and their covariance matrix is
\begin{equation}\label{eqn:case.iff.cond}
    \left[
    \begin{array}{cc} \sigma^2_1 \sigma_X^2 & \sigma^2_1 \sigma_X^2\\ \sigma^2_1 \sigma_X^2 & \sigma^2_2 \sigma_X^2
    \end{array} \right].
\end{equation}

\begin{proof}
Refer to Appendix~\ref{appendix:proof.theorem:case.independence}.
\end{proof}
\end{theorem}

The following theorem states that $\NoiseS$ and $\NoiseM-\NoiseS$
being independent is a sufficient condition for
Equation~\ref{eqn:case.goal} to hold.
\begin{theorem}\label{theorem:case.suff.cond}
Given that $\NoiseS \sim N(0, \sigma^2_1 \sigma_X^2)$ and $\NoiseM \sim N(0,
\sigma^2_2 \sigma_X^2)$, and $\sigma_1^2<\sigma^2_2$, if $\NoiseS$ and $\NoiseM-\NoiseS$ are independent, then Equation~\ref{eqn:case.goal} holds.

\begin{proof}
Refer to Appendix~\ref{appendix:proof.theorem:case.suff.cond}.
\end{proof}

\textbf{Example}. We now revisit the example in Section~\ref{ssec:case.ind.noise} to show that collusion does not improve estimation accuracy in our scheme. Assume that $\noisyDBS$ and $\noisyDBM$ are generated following the proposed solution, i.e.,
$\NoiseS$ and $\NoiseM$ are jointly Gaussian and their covariance matrix is
$\left[\begin{array}{cc} 1 & 1\\ 1 & 4
    \end{array} \right]$.
The reconstructed data after the collusion between Alice and Bob using Eqn. (\ref{eqn:LLSE.yy})
are $\hat{\db}(\noisyDBS) = [9.5, 10.5]^T + 0.5\NoiseS$. The average estimation error is
$$\frac{1}{2} E[(\hat{\db} - \db)^T (\hat{\db} - \db)]
= 0.125 E[\NoiseS^T \NoiseS] + 0.25 = 0.5.$$
This error of joint estimation is the same as the error of estimation using only the least perturbed copy. Thus the collusion does not result in a smaller error in our scheme.  \hfill $\Box$


\begin{remark}
Intuitively, since $\noisyDBM$ is a perturbed observation of $\noisyDBS$ as
shown in Equation \ref{eqn:z2minusz1}, $\noisyDBM$ cannot provide extra innovative information to improve the
estimation accuracy achieved by utilizing only $\noisyDBS$, and Equation~\ref{eqn:case.goal} is satisfied.

This sufficient condition is key in achieving the privacy goal in
this simple case, as well as in the general cases, on
which we elaborate in Section~\ref{sec:solution}.
\end{remark}
\end{theorem}

Following the above analysis, our solution to
this simple case is as follows:
\begin{itemize}
  \item Given $\varianceS$ and $\varianceM$, construct the covariance matrix of $Z_1$ and $Z_2$ as in Equation~\ref{eqn:case.iff.cond}. Derive the joint distribution of $\NoiseS$ and $\NoiseM$.
  \item Compute the conditional distribution of $\NoiseS$ given $\NoiseM$. Generate $\NoiseS$ according to this conditional distribution.
  \item Generate the desired $\noisyDBS=\db+\NoiseS$.
\end{itemize}
In this way, $\NoiseS$ and $\NoiseM-\NoiseS$ are guaranteed to be independent; hence,
Equation~\ref{eqn:case.goal} is satisfied.


\section{Solution to General Cases}\label{sec:solution}
We now show that the solutions to the
general cases of arbitrarily fine trust levels follow naturally from
that to the two trust level case studied in
Section~\ref{sec:case.study}.

\subsection{Shaping the Noise}
\subsubsection{Independent Noise Revisited} \label{sec:ind.noise}
In Section~\ref{sec:case.study}, we show that adding independent
noise to generate two differently perturbed copies, although
convenient, fails to achieve our privacy goal. The increase in the
number of independently generated copies aggravates the situation;
the estimation error actually goes to zero as this number increases
indefinitely. In turn, the attackers can perfectly reconstruct the
original data. We formalize this observation in the following
theorem.

\begin{theorem}\label{theorem:ind.noise}
Let $\yy=[Y_1^T,\ldots, Y^T_M]^T$ be a vector containing $M$ perturbed copies. Assume that $\yy$ is generated from the
original data $\db$ as follows:
\[
    \yy = HX + \zz,
\]
where $H =\left[I_N, \ldots, I_N\right]^T$,
and $\zz = [\Noise^T_1, \ldots, \Noise^T_M]^T$ with $Z_i \sim N(0,
\sigma^2_{Z_i}K_X)$ is the noise vector.

If noise $Z_i, 1\leq i\leq \levels$ are mutually independent, then the
square errors between the LLSE estimate $\db$ and $\hat{X}(\yy)$ are
the diagonal terms of the following matrix
\[
    \left(1+\sum_{i=1}^{M}\frac{1}{\sigma^2_{Z_i}} \right)^{-1} K_X.
\]
As $\levels$ increases, the estimation errors decrease, so does the distortion $\dis(\db, \hat{X}(\yy))$.

\begin{proof}
Refer to Appendix~\ref{appendix:proof.theorem:ind.noise}.
\end{proof}

\begin{remark}
The theorem says that when adding a new copy that is perturbed by independent noise, the estimation error decreases.
It agrees with the intuition that a new independently-perturbed copy adds extra innovative information to improve
the estimation accuracy.
\end{remark}
\end{theorem}

We conclude that noise $Z_i, 1\leq i\leq \levels$ should not be
generated independently.

\subsubsection{Properly Correlated Noise}
We show by the case study that the key to achieving the desired
privacy goal is to have noise $Z_i, 1\leq i\leq \levels$
properly correlated. To this end, we further develop the pattern
found in the $2 \times 2$ noise covariance matrix in
Equation~\ref{eqn:case.iff.cond} into a {\em corner-wave} property
for a multi-dimensional noise covariance matrix. This property
becomes the cornerstone of Theorem~\ref{theorem:suff.cond} which is
a generalization of Theorem~\ref{theorem:case.independence} and
~\ref{theorem:case.suff.cond}.\\

{\noindent\bf Corner-wave Property} Theorem~\ref{theorem:suff.cond}
states that for $\levels$ perturbed copies, the privacy goal in
Equation~\ref{eqn:goal} is achieved if the noise
covariance matrix $K_{\zz}$ has the corner-wave pattern as shown in
Equation~\ref{eqn:suff.cond}. Specifically, we say that an $\levels
\times \levels$ square matrix has the corner-wave property if, for
every $i$ from $1$ to $M$, the following entries have the same value
as the $(i,i)^{th}$ entry:
\begin{itemize}
  \item all entries to the right of the $(i,i)^{th}$ entry in row
  $i$,
  \item all entries below the $(i,i)^{th}$ entry in column $i$.
\end{itemize}
The distribution of the entries in such a matrix looks like corner-waves
originated from the lower right corner.

\begin{theorem}\label{theorem:suff.cond}
Let $\yy=[Y_1^T,\ldots, Y^T_M]^T$ represent an arbitrary number
of perturbed copies. Assume that $\yy$ is generated from the
original data $\db$ as follows:
\[
    \yy = HX + \zz,
\]
where $H =\left[I_N, \ldots, I_N\right]^T$,
and $\zz = [\Noise^T_1, \ldots, \Noise^T_M]^T$ with $Z_i
\sim N(0, \sigma^2_{Z_i}K_X)$ is the noise vector. Without loss
of generality, we further assume
\begin{equation}\label{eqn:asmp:sigma_z}
   \sigma^2_{Z_i}< \sigma^2_{Z_{i+1}},\;\;\; \forall i=1,\ldots, \levels-1.
\end{equation}
Then the following equation holds
\[
    \dis(\db, \hat{\db}(\yy)) = \min_{i=1,\ldots, \levels}\dis(\db, \hat{\db}(Y_i)) = \frac{\sigma_{Z_1}^2}{\sigma_{Z_1}^2+1}\frac{1}{N}Tr(K_X),
\]
if $\zz$ is a jointly Gaussian vector and its covariance matrix $K_{\zz}$ is given by
\begin{eqnarray}\label{eqn:suff.cond}
    K_{\zz} &=&
\left[\begin{array}{cccc}
\sigma_{Z_{1}}^{2}K_{X} & \sigma_{Z_{1}}^{2}K_{X} & \cdots & \sigma_{Z_{1}}^{2}K_{X}\\
\sigma_{Z_{1}}^{2}K_{X} & \sigma_{Z_{2}}^{2}K_{X} & \cdots & \sigma_{Z_{2}}^{2}K_{X}\\
\vdots & \vdots & \ddots & \vdots\\
\sigma_{Z_{1}}^{2}K_{X} & \sigma_{Z_{2}}^{2}K_{X} & \cdots & \sigma_{Z_{M}}^{2}K_{X}\end{array}\right].
\end{eqnarray}

\begin{proof}
Refer to Appendix~\ref{appendix:proof.theorem:suff.cond}.
\end{proof}

\begin{remark}
The corner-wave property of $K_{\zz}$ given in Equation \ref{eqn:suff.cond} guarantees that Equation \ref{eqn:goal} holds.
Therefore, the diversity attack does not help to improve the estimation accuracy.
\end{remark}

\end{theorem}

Moreover, for any subset of these $\levels$ perturbed copies, the
covariance matrix of the corresponding noise also has the
corner-wave property, and thus the privacy goal is achieved. We
summarize this observation in Corollary~\ref{cor:suff.cond}.

\begin{cor}\label{cor:suff.cond}
If the privacy goal in Equation~\ref{eqn:goal} is achieved with respect to
$\levels$ perturbed data $Y_1, \ldots, Y_M$, then the goal is
also achieved with respect to any subset of $\{Y_1, \ldots, Y_M\}$.
\end{cor}

Based on Theorem~\ref{theorem:suff.cond} and
Corollary~\ref{cor:suff.cond}, one way to achieve the privacy goal
in Equation~\ref{eqn:goal} is to ensure that noise $\zz$
is a jointly Gaussian vector and follows $N(0, K_{\zz})$ where
$K_{\zz}$ is given by Equation~\ref{eqn:suff.cond}. We
consider two scenarios when generating noise $\zz$ and the
corresponding perturbed copies $\yy$. We discuss these two
scenarios in the following two sections.

\subsection{Batch Generation}\label{ssec:all-in-once}
In the first scenario, the data owner determines the
$\levels$ trust levels \emph{a priori}, and generates $\levels$ perturbed
copies of the data in one batch. In this case, all trust levels are
predefined and $\sigma^2_{Z_1}$ to $\sigma^2_{Z_M}$ are given when
generating the noise. We refer to this scenario as the {\em batch
generation}.

We propose two batch algorithms. Algorithm 1 generates noise
$\NoiseS$ to $\Noise_{\levels}$ in parallel while Algorithm 2
sequentially.

\subsubsection{Algorithm 1: Parallel Generation}
Without loss of generality, we assume
$\sigma^2_{Z_{i}}<\sigma^2_{Z_{i+1}}$ where $1\leq i\leq M-1$.
Algorithm~\ref{alg:A1} generates the components of noise $\zz$, i.e.,
$\NoiseS$ to $\Noise_{\levels}$, simultaneously based on the
following probability distribution function, for any real $(N\cdot
M)$-dimension vector $v$,
\begin{equation}\label{eqn:zz.pdf}
    f_{\zz}(v) = \frac{1}{\sqrt{(2\pi)^M \det (K_{\zz})}}e^{-\frac{1}{2}v^TK^{-1}_{\zz}v},
\end{equation}
where $K_{\zz}$ is given by Equation~\ref{eqn:suff.cond}.

Algorithm~\ref{alg:A1} then constructs $\yy$ as $HX+\zz$
and outputs it. We refer to Algorithm~\ref{alg:A1} as {\em parallel
generation}.
\begin{algorithm}[hbt]
\begin{algorithmic}[1]
    \STATE{// Input: $\db$, $K_{\db}$, and $\sigma^2_{Z_1}$ to $\sigma^2_{Z_M}$}
    \STATE{// Output: $\yy$} 
    \STATE{Construct $K_{\zz}$ with $K_{\db}$ and $\sigma^2_{Z_1}$ to $\sigma^2_{Z_M}$, according to Equation~\ref{eqn:suff.cond}}
    \STATE{Generate $\zz$ 
    with $K_{\zz}$, according to Equation~\ref{eqn:zz.pdf}}
    \STATE{Generate $\yy = H\db +\zz$}
    \STATE{Output $\yy$}
    \caption{: Parallel Generation
    }\label{alg:A1}
\end{algorithmic}
\end{algorithm}

Algorithm~\ref{alg:A1} serves as a baseline algorithm for
the next two algorithms.

\subsubsection{Algorithm 2: Sequential Generation}
The large memory requirement of Algorithm 1 motivates us
to seek for a memory efficient solution. Instead of parallel
generation, sequentially generating noise $Z_1$ to $Z_M$, each of
which a Gaussian vector of $N$ dimension. The validity of the alternative procedure is based
on the insight in the following theorem.
\begin{theorem}\label{theorem:independence}
Consider $\zz = [\Noise^T_1, \ldots, \Noise^T_M]^T$ where $Z_i \sim N(0, K_{Z_i})$ with $K_{Z_i} = \sigma^2_{Z_i}K_X$. Without loss of generality, further assume
   $$\sigma^2_{Z_i}< \sigma^2_{Z_{i+1}},\;\;\; \forall i=1,\ldots, \levels-1.$$
Then $\zz$ is a jointly Gaussian vector and $K_{\zz}$ has the form in Equation~\ref{eqn:suff.cond}, if and only if $Z_1$, and $(Z_{i}-Z_{i-1}), i = 2, ..., M$ are mutually independent.

\begin{proof}
Refer to Appendix~\ref{appendix:proof.theorem:independence}.
\end{proof}
\end{theorem}

Based on Theorem~\ref{theorem:independence}, Algorithm 2
sequentially generates $M$ independent noise $Z_1$, and
$(Z_{i}-Z_{i-1})$ for $i$ from $2$ to $M$. Noise $\Noise_i$ is then
simply $(\Noise_i-\Noise_{i-1}) + \Noise_{i-1}$ for $i$ from 2 to
$\levels$. Finally Algorithms 2 generates the perturbed copies $Y_1$
to $Y_M$ by adding the corresponding noise. We refer to Algorithm 2
as {\em sequential generation}.

\begin{algorithm}[hbt]
\begin{algorithmic}[1]
    \STATE{// Input: $\db$, $K_{\db}$, and $\sigma^2_{Z_1}$ to $\sigma^2_{Z_M}$}
    \STATE{// Output: $Y_1$ to $Y_M$}
    \STATE{Construct $Z_1 \sim N(0, \sigma^2_{Z_1}K_{\db})$} 
    \STATE{Generate $Y_1=X+Z_1$}
    \STATE{Output $Y_1$}
    \FOR {$i$ from $2$ to $\levels$}
        \STATE {Construct noise $\xi \sim N(0, (\sigma^2_{Z_{i}}-\sigma^2_{Z_{i-1}})K_{\db})$} 
        \STATE{Generate $Y_{i}=Y_{i-1}+\xi$}
        \STATE{Output $Y_i$}
    \ENDFOR
    \caption{: Sequential Generation
    }\label{alg:A2}
\end{algorithmic}
\end{algorithm}

We now explain intuitively why the mutual independence requirement
for $Z_1$, and $(Z_{i}-Z_{i-1})$ for $i$ from $2$ to $M$ is
sufficient to achieve our privacy goal in Equation~\ref{eqn:goal}.

We rewrite $Y_i$ as $X+Z_1+\sum_{j=2}^i (Z_j - Z_{j-1})$. Since
$\db$, $Z_1$ and $Z_j-Z_{j-1}$ for $j=2,\ldots, M$ are mutually
independent, $Y_i, 2\leq i\leq \levels$ are perturbed observations
of $Y_1$. Intuitively all information in them that are useful for
estimating $\db$ is inherited from $Y_1$. As such, given $Y_1$,
$Y_i, 2\leq i\leq \levels$ provides no extra innovative information
to improve the estimation accuracy. Similar analysis applies to any
subset of $Y_1$ to $Y_M$. Hence, Equation~\ref{eqn:goal} is
satisfied. This intuition is similar to the explanation for the case
study in Section~\ref{sec:case.study}.

\subsubsection{Disadvantages}
The main disadvantage of the batch generation approach is
that it requires a data owner to foresee all possible trust levels \emph{a
priori}.

This obligatory requirement is not flexible and sometimes impossible
to meet. One such scenario for the latter arises in our case study.
After the data owner already released a perturbed copy $Y_2$, a new
request for a less distorted copy $Y_1$ arrives. The
sequential generation algorithm cannot handle such requests since
the trust level of the new request is lower than the existing one.
In today's ever changing world, it is desirable to have technologies
that adapt to the dynamics of the society. In our problem setting,
generating new perturbed copies on-demand would be a desirable
feature.

\subsection{On Demand Generation} \label{ssec:on.the.fly}
As opposed to the batch generation, new perturbed copies are
introduced on demand in this second scenario. Since the requests may
be arbitrary, the trust levels corresponding to the new copies would
be arbitrary as well. The new copies can be either lower or higher
than the existing trust levels. We refer this scenario as {\em
on-demand} generation. Achieving the privacy goal in this scenario
will give data owners the maximum flexibility in providing MLT-PPDM
services.


We assume $L(L<M)$ existing copies of $Y_1$ to $Y_{L}$. We
also assume that the data owner, upon requests, generates additional
$M-L$ copies of $Y_{L+1}$ to $Y_M$. Thus there will be $M$ copies in
total. Note in this subsection $\sigma^2_{Z_1}$ to $\sigma^2_{Z_M}$
can be in any order. Finally, we define vectors $\zz'$ and $\zz''$
as
\[
    \zz' =\left[\begin{array}{c}
Z_{1}\\
\vdots\\
Z_{L}\end{array}\right]\mbox{  and  }    \zz'' =\left[\begin{array}{c}
Z_{L+1}\\
\vdots\\
Z_{M}\end{array}\right].
\]


According to Theorem~\ref{theorem:suff.cond}, the data owner should generate new noise $\zz''$ in such a way that the covariance matrix of $\zz=[{\zz'}^T {\zz''}^T]^T$ has corner-wave property, and they are jointly Gaussian.

The desired covariance matrix $K_{\zz}$ can be constructed according to Equation~\ref{eqn:suff.cond} (after  properly ordering $Z_{1}$ to $Z_M$ according to $\sigma^2_{Z_1}$ to $\sigma^2_{Z_M}$).

According to Section \ref{ssec:jnt.gaussian}, it is sufficient and necessary for the conditional distribution of $\zz''$ given that $\zz'$ takes any value $v_1$ to be a Gaussian with mean
\begin{equation}\label{eqn:zz''.cond.expt}
    K_{\zz''\zz'}K_{\zz'}^{-1}v_1
\end{equation}
and covariance
\begin{equation}\label{eqn:zz''.cond.var}
    K_{\zz''} - K_{\zz''\zz'}K_{\zz'}^{-1}K_{\zz''\zz'}^T,
\end{equation}
where $K_{\zz'}$ is the covariance matrix of $\zz'$, $K_{\zz''\zz'}$ is the desired covariance matrix between $\zz''$ and $\zz'$, and $K_{\zz''}$ is the desired covariance matrix of $\zz''$.

Note $K_{\zz'}$ is known to the data owner, and $K_{\zz''\zz'}$ and $K_{\zz''}$ can be extracted from the desired covariance matrix $K_{\zz}$. We turn the above analysis into Algorithm~\ref{alg:A3}.
\begin{algorithm}[hbt]
\begin{algorithmic}[1]
    \STATE{// Input: $\db$, $K_{\db}$, $\sigma^2_{Z_1}$ to $\sigma^2_{Z_M}$, and  values of $\zz'$: $v_1$}
    \STATE{// Output: New copies $\zz''$}
    \STATE{Construct $K_{\zz}$ with $K_{\db}$ and $\sigma^2_{Z_1}$ to $\sigma^2_{Z_M}$, according to Equation~\ref{eqn:suff.cond}}
    \STATE{Extract $K_{\zz'}$, $K_{\zz''\zz'}$, and $K_{\zz''}$ from $K_{\zz}$}
    \STATE{Generate $\zz''$ as a Gaussian with mean and variance in Equation~\ref{eqn:zz''.cond.expt} and~\ref{eqn:zz''.cond.var}, respectively}
    \FOR {$i$ from $L+1$ to $\levels$}
        \STATE{Generate $Y_{i}=X+Z_i$}
        \STATE{Output $Y_i$}
    \ENDFOR
    \caption{: On Demand Generation}\label{alg:A3}
\end{algorithmic}
\end{algorithm}

\subsection{Time and Space Complexity}\label{ssec:Space-Time-Complex}
In this subsection, we study the time and space complexity of the three algorithms.
One may notice that all the covariance matrices of noise in the three algorithms, such as Equation \ref{eqn:suff.cond} and Equation \ref{eqn:zz''.cond.var}, can be written as the Kronecker product of two matrices. For such covariance matrices, we have the following
observation:
\begin{lemma} \label{lem:jointGauss}
Assume that $\mu$ and $K$ are the mean and covariance matrix of the jointly Gaussian random vector $\G$. If $K_{\G} =
\Sigma_{\G} \otimes K_0$, where $\Sigma_{\G}$ and $K_0$ are $P \times P$ and $Q \times Q$, respectively, and $K_0$ is also a covariance matrix, then the time complexity of generating $\G$ is $O(P^3 + Q^3)$. 
\end{lemma}

\begin{proof}
Refer to Appendix \ref{appendix:complexity:lemma}.
\end{proof}

\begin{remark}
Directly generating $\G$ using $K_\G$, the complexity is $O(P^3 Q^3)$. Viewing $K_\G$ as a Kronecker product of two matrices of smaller dimensions, we can utilize the properties of Kronecker product to reduce the complexity to $O(P^3 + Q^3)$. 

The proof suggests an efficient implementation of the proposed three
algorithms. Note that for each algorithm, the time complexity may be
further reduced.
\end{remark}

Utilizing Lemma \ref{lem:jointGauss}, we give the following theorems on the time and space complexity of the proposed three algorithms.
\begin{theorem} \label{theorem:complexity:alg1}
Given an $N$-dimensional data vector $X$, the time complexity of generating $M$ perturbed copies using Algorithm~\ref{alg:A1} is $O(N^3 + M N^2)$, and the space complexity is $O(M + N^2)$.
\end{theorem}

\begin{proof}
Refer to Appendix \ref{appendix:complexity:alg1}.
\end{proof}

\begin{theorem} \label{theorem:complexity:alg2}
Given an $N$-dimensional data vector $X$, the time complexity of generating $M$ perturbed copies using Algorithm~\ref{alg:A2} is
$O(N^3+M N^2)$, and the space complexity is $O(N^2)$.
\end{theorem}

\begin{proof}
Refer to Appendix \ref{appendix:complexity:alg2}.
\end{proof}

\begin{remark}
Using a similar set of arguments, we can show the time complexity of the independent noise scheme described in Section \ref{sec:ind.noise} is the same as Algorithm 2.
\end{remark}

\begin{table*}[t]
  \centering
    \caption{Comparison of applicabilities, space complexity, and time complexity of three proposed algorithms.}\label{tab:algs.comps}
    \begin{tabular}{c|c|c|c|c}
    \hline
     & Batch  & On-demand & Space & Time \tabularnewline
     &  Generation           &   Generation        & Complexity & Complexity\tabularnewline
    \hline
    \hline
    Algorithm 1 & \checkmark &  & $O(M + N^2)$ & $O(N^3 + M N^2)$\tabularnewline
    \hline
    Algorithm 2 & \checkmark &  & $O(N^2)$ & $O(N^3 + M N^2)$\tabularnewline
    \hline
    Algorithm 3 & \checkmark & \checkmark & $O(M^2 + N^2)$ & $O(M^3 + N^3)$ \tabularnewline
    \hline
    \end{tabular}
\end{table*}

\begin{theorem} \label{theorem:complexity:alg3}
Given an $N$-dimensional data vector $X$ and $L$ ($1 \leq L \leq M-1$) perturbed copies of $X$, the time complexity of generating $(M-L)$ perturbed copies using Algorithm~\ref{alg:A3} is $O(M^3 + N^3)$, and the space complexity is $O(M^2 + N^2)$.
\end{theorem}

\begin{proof}
Refer to Appendix \ref{appendix:complexity:alg3}.
\end{proof}


Table~\ref{tab:algs.comps} compares the applicabilities and complexity of the three proposed algorithms. In summary, Algorithms~\ref{alg:A1} and~\ref{alg:A2} have less space and time complexity than Algorithm \ref{alg:A3}; Algorithm~\ref{alg:A3} offers data owners maximum flexibility by generating perturbed copies in an on-demand fashion.

\section{Experiments}\label{sec:simulation}

\subsection{Methodology and Settings}\label{ssec:sim.methodology}
We design two experiments, performance test (Experiment 1) and scalability test (Experiment 2).
Experiment 1 explores answers to the following questions numerically:
\begin{itemize}
  \item How severe can LLSE-based diversity attacks be, given that the perturbed copies at different trust levels are generated independently?
  \item How effective is our proposed scheme against LLSE-based diversity attacks, compared to the above independent noise scheme?
  \item How does an adversary's knowledge affect the power of such attacks?
\end{itemize}
Experiment 2 demonstrates the runtime of our proposed Algorithm 3.

\begin{figure*}[thb]
  \centering
    \begin{tabular}{c@{\extracolsep{1.0em}}c}
    \includegraphics[width=0.75\columnwidth]{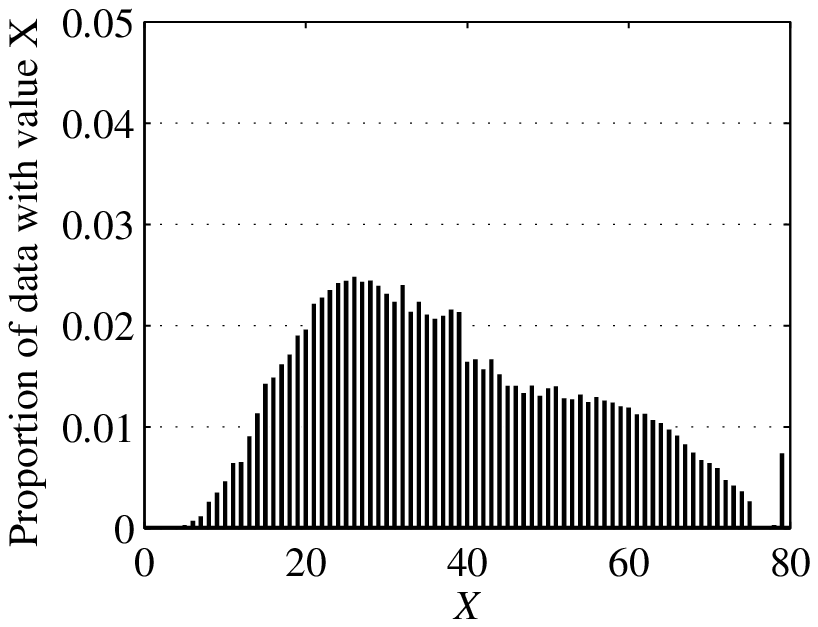} &
    \includegraphics[width=0.75\columnwidth]{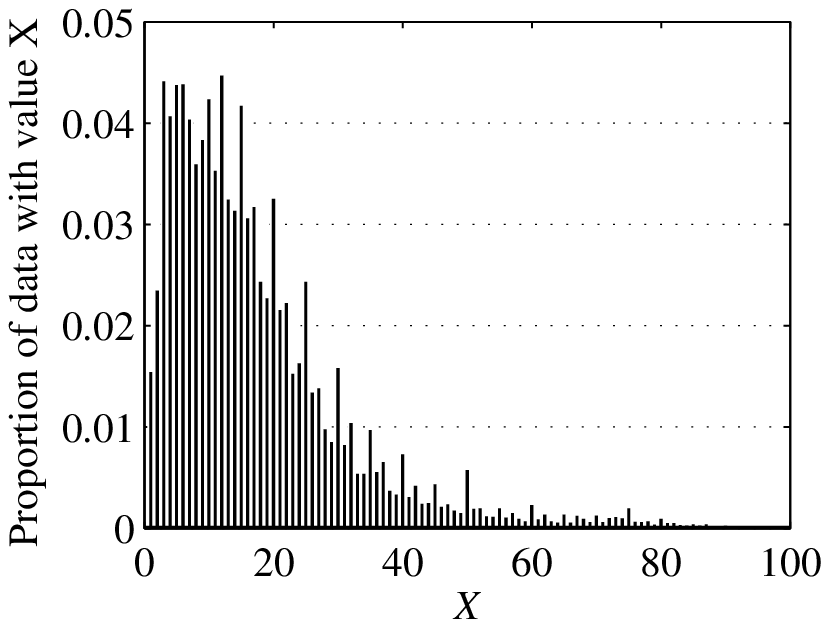} \\
    (a) \emph{Age} & (b) \emph{Income}
    \end{tabular}
  \caption{Distribution of sensitive values \emph{Age} and \emph{Income}.}\label{fig:expr.data}
\end{figure*}

We run our experiments on a real dataset CENSUS \cite{mpc.data}, which is commonly used in the literature of privacy preservation such as \cite{xiao2008output}, for carrying out the experiments and evaluating their performance in a fully controlled manner. This dataset contains one million tuples with four attributes: \emph{Age}, \emph{Education}, \emph{Occupation}, and \emph{Income}. 
We take the first $10^5$ tuples and conduct the experiments on the \emph{Age} and \emph{Income} attributes. The statistics and distribution of the data are shown in Table \ref{tab:expr.data} and Figure \ref{fig:expr.data}, respectively.

\begin{table}[b]
  \centering
    \caption{Statistics of the original data \emph{Age} and \emph{Income}.}\label{tab:expr.data}
    \begin{tabular}{c|c|c}
    \hline
     & Mean $\mu_X$ & Variance $\sigma_X^2$ \\
    \hline
    \hline
    \emph{Age} & $50.06$ & $303.03$ \\
    \hline
    \emph{Income} & $16.57$ & $219.92$ \\
    \hline
    \end{tabular}
\end{table}

Given data $X$ (\emph{Age} and \emph{Income}), to generate perturbed copies $Y_i$ at different trust levels $i$, we
generate Gaussian noise $Z_i$ according to $N(0,
\sigma^2_{Z_i}K_X)$, and add $Z_i$ to $X$. The constant
$\sigma^2_{Z_i}$ represents the perturbation magnitude determined by the
data owner according to the trust level $i$. The noise for different
trust levels are generated either independently, or in a properly
correlated manner following our proposed solution in
Section~\ref{sec:solution}.

Data miners can access one or more perturbed copies $Y_i$, either according to application scenario setting or by collusion among themselves. Recall our assumption that
data miners perform joint LLSE estimation to reconstruct $X$.
We study two classes of data miners with different knowledge about the original data and noise:
\begin{itemize}
  \item the first class of adversaries has perfect knowledge, i.e., the exact values of $\mu_X$, $K_X$ and $\sigma^2_{Z_i}$ for every trust level $i$;
  \item  the second class of adversaries has partial knowledge, i.e., the exact values of $\sigma^2_{Z_i}$ for every trust level $i$, but not $\mu_X$ and $K_X$.
\end{itemize}

To perform LLSE estimation, data miners with partial knowledge
estimate $\mu_X$ and $K_X$ using their perturbed copies. For each
$Y_i$, its mean is simply $\mu_X$, and its covariance matrix is
$(1+\sigma^2_{Z_i})K_X$. Knowing the exact values of
$\sigma^2_{Z_i}$, a data miner can estimate $\mu_X$ and $K_X$ using
the sample mean and sample covariance matrix of $Y_i$. Accuracy of
such estimation depends on the sample size; the larger the sample size, the more accurate the estimation of $\mu_X$ and $K_X$.

In Experiment 1, we use two performance metrics, average normalized estimation error and distribution of estimation error.
For LLSE estimate of $X$ based on $\yy$, i.e., $\hat{X}(\yy)$, we define its normalized estimation error as
\[
    \frac{\dis(X, \hat{X}(\yy))}{Tr(K_{X})}.
\]
It takes values between $0$ and $1$. The smaller it is, the more accurate the LLSE estimation is. It generally decreases as more perturbed copies are used in the LLSE estimation. When showing the distribution of the estimation error, we use $\sqrt{\dis(X, \hat{X}(\yy))}$ directly, and one may see how large the distortion is, compared to the values of the original data shown in Fig. \ref{fig:expr.data}, as we do not normalize it.
The distribution is represented by a histogram as well as a cumulative histogram. The curve of cumulative histogram starts from 0 and increases to 1. The faster the curve approaches 1, i.e., the bigger proportion of accurate estimates, the better the LLSE-based diversity attack performs.
We conduct experiments on data with two attributes (i.e., $N=2$); however, for ease of illustration, we show the performance on different attributes separately.

\begin{figure*}[thb]
  \centering
    \begin{tabular}{c@{\extracolsep{1.0em}}c}
    \includegraphics[width=0.96\columnwidth]{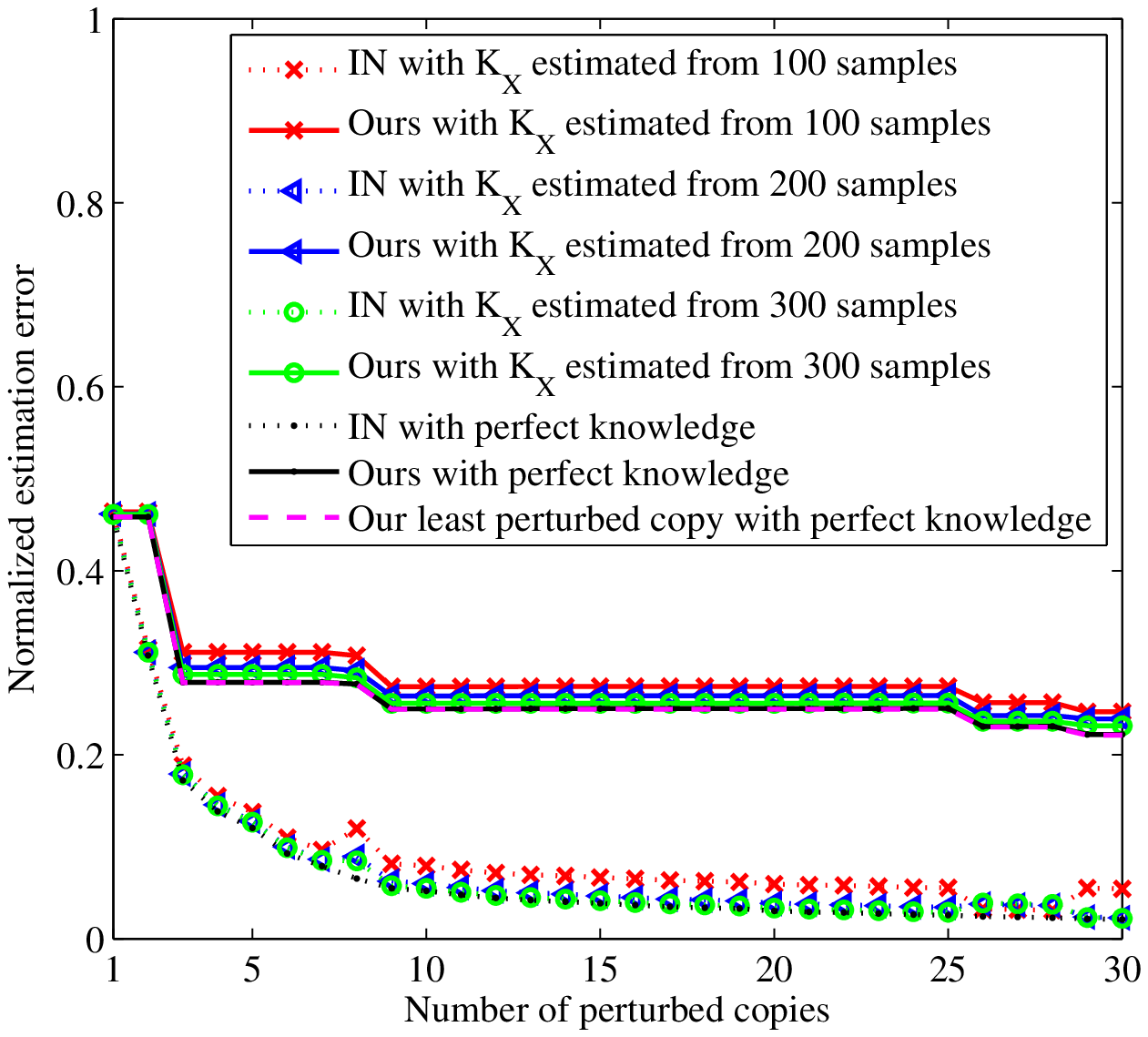} &
    \includegraphics[width=0.96\columnwidth]{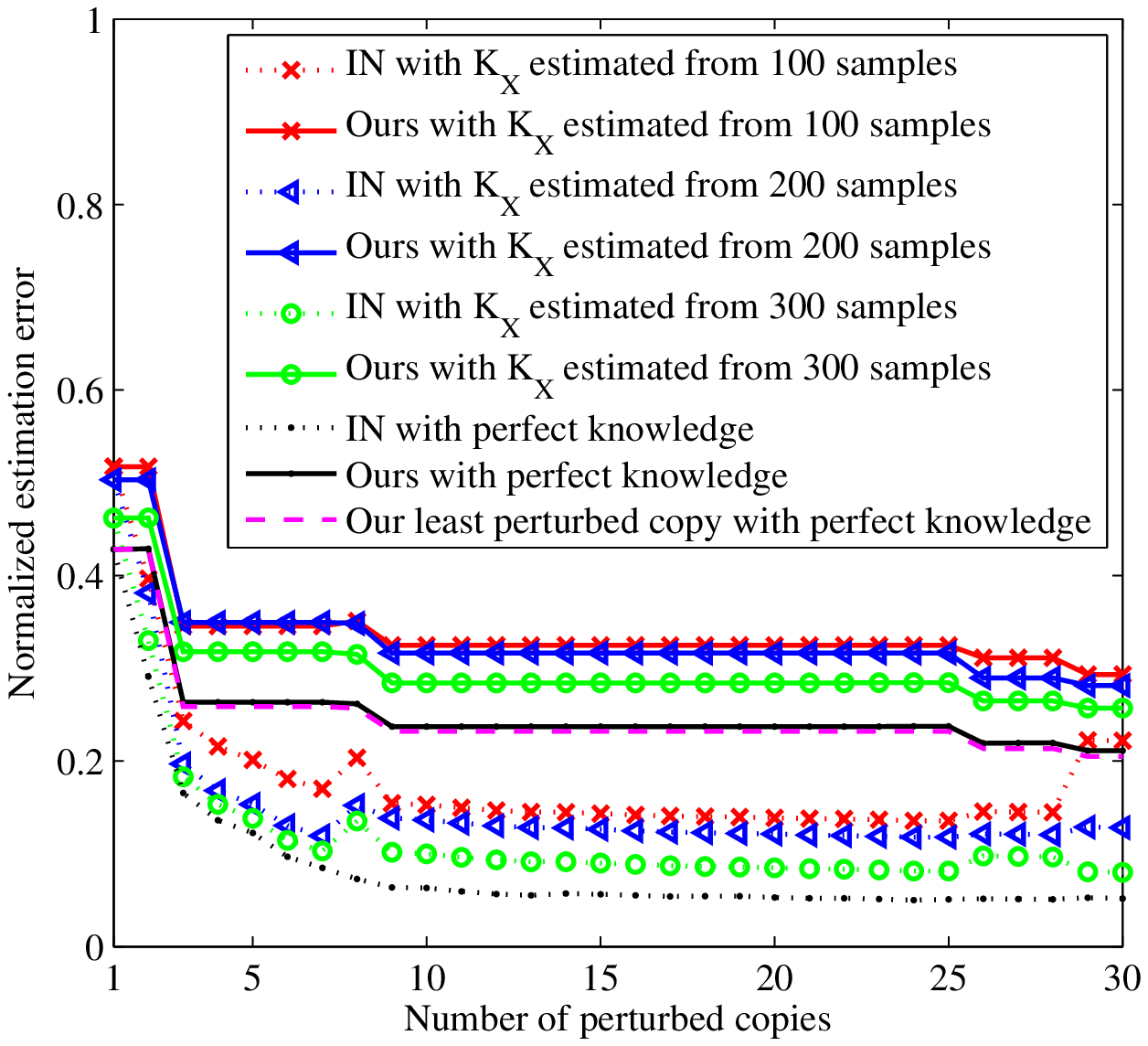} \\
    (a) \emph{Age} &
    (b) \emph{Income} \\
    \multicolumn{2}{c}{
    \includegraphics[width=1.26\columnwidth]{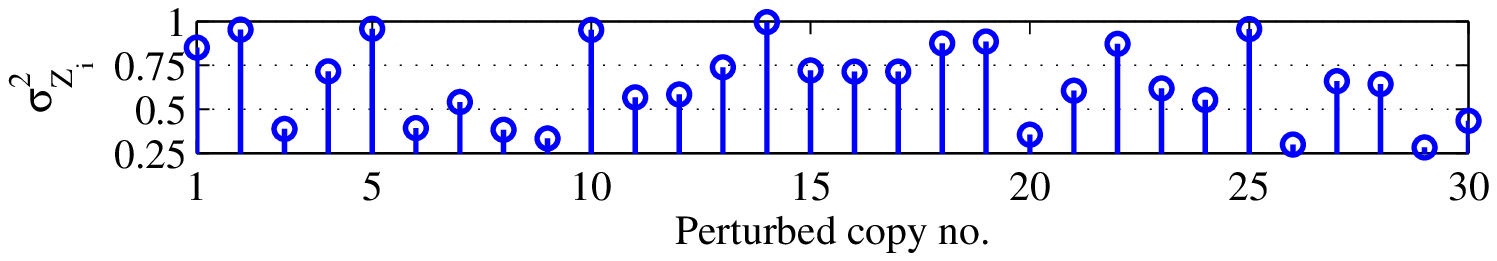}} \\
    \end{tabular}
  \caption{Comparisons of average normalized estimation error of the independent noise scheme (denoted as \emph{IN}) and our scheme (denoted as \emph{Ours}) on the data \emph{Age} (a) and \emph{Income} (b), respectively. The average normalized estimation error of each setting is shown as a function of the number of generated perturbed copies. Note that using our algorithm, the curve of attacks utilizing the least perturbed copy overlaps with the curve of attacks utilizing all the available $M$ copies. Perturbation magnitude $\sigma^2_{Z_i}$ is shown as a function of perturbed copy number $i$ at the bottom.}\label{fig:nee-compare}
\end{figure*}
\subsection{Experiment 1: Performance Test}
In this subsection, we show the superiority of our scheme over the scheme that simply adds independent noise,
and how data miner's knowledge affects the power of LLSE-based diversity attacks. Algorithm \ref{alg:A3} is used for
the experiment due to its maximum flexibility among the three proposed algorithms.

$M$ perturbed copies $Y_i$, $1\leq i\leq M$, are generated one by one upon requests, adding independent noise to the original data or
using our proposed Algorithm \ref{alg:A3}. Each
request is at a different trust level with corresponding
$\sigma^2_{Z_i}$ randomly generated in $[0.25, 1]$.
Figure~\ref{fig:nee-compare} shows $\sigma^2_{Z_i}$ as a
function of perturbed copy number $i$.

We assume that data miners can access all the $M$ perturbed copies. This setting represents
the most severe attack scenario where data miners jointly estimate $X$ using all the available $M$
perturbed copies. Since the perturbed copies are released one by one, the number of the available perturbed
copies also increases one by one.

We also assume that data miners with partial knowledge estimate $\mu_X$
and $K_X$ with different sample sizes. In particular, we assume that
they have $100 N^2$, $200 N^2$ and $300 N^2$ samples, where $N^2$ is
the number of entries in $K_X$ and $N=2$ in our experiments.

Figures~\ref{fig:nee-compare}(a) and (b) show the normalized estimation errors of both schemes as
a function of the number of perturbed copies, on attributes \emph{Age} and \emph{Income}, respectively.

The results of the experiments clearly show that the diversity gain in joint estimation reduces the normalized estimation error dramatically. While for our algorithm, we find that the estimation error
drops only when a perturbed copy with minimum perturbation magnitude
so far becomes available. Using our algorithm, the curve of attacks utilizing the least perturbed copy overlaps with the curve of attacks utilizing
all the available $M$ copies. The above observations imply that the joint
estimation based on all existing copies is only as good as the
estimation based on the copy with the minimum privacy, and there is no diversity gain in performing the LLSE estimation jointly. Moreover, we have verified that the estimation error matches our analytical result in Theorem~\ref{theorem:suff.cond}.

We also find that when data miners have
perfect knowledge, the normalized estimation error decreases
monotonically as $M$ increases for copies perturbed by independent noise. This trend indicates a perfect
reconstruction of $X$ when $M$ goes to infinity. It also confirms Theorem~\ref{theorem:ind.noise} empirically.

On the other hand, if the adversaries have to estimate $\mu_X$ and $K_X$ from samples, i.e., the attackers have partial
knowledge, the curve flattens and even slightly increases as $M$
becomes large. This is because the estimation error depends not only on
the number of perturbed copies, but also on the precision of $\mu_X$
and $K_X$. The estimation based on inaccurately estimated $m_X$
and $K_X$ is not optimal. Consequently, the estimation
accuracy does not always improve as $M$ increases.
Figure~\ref{fig:nee-compare} also shows that adversaries having more
samples perform better in estimating $\mu_X$ and $K_X$, resulting in
improved overall accuracy.


\begin{figure*}[thb]
  \centering
    \begin{tabular}{c@{\extracolsep{3.0em}}c}
    \includegraphics[width=0.93\columnwidth]{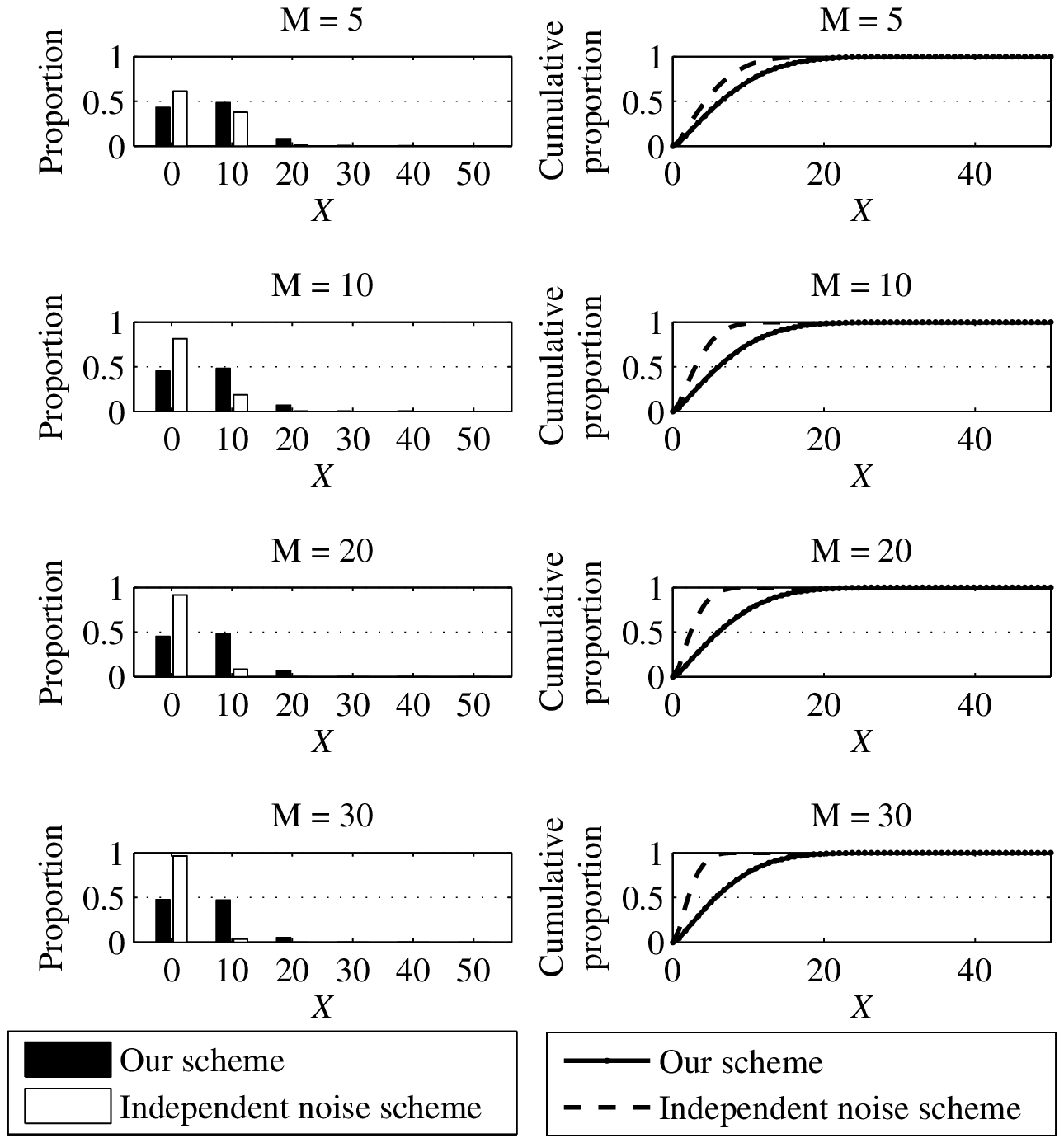} &
    \includegraphics[width=0.93\columnwidth]{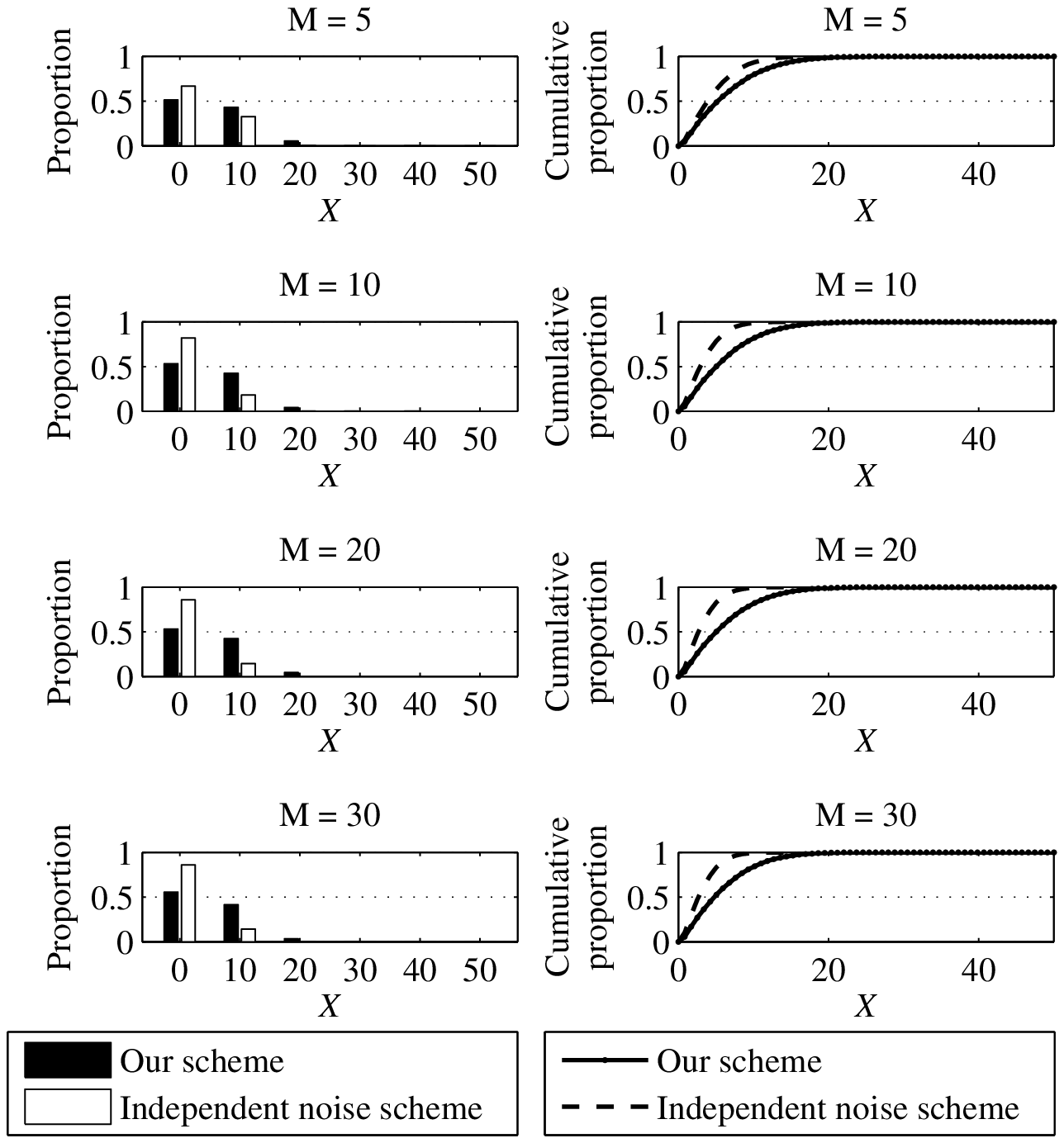} \\
    (a) \emph{Age} &
    (b) \emph{Income} \\
    \end{tabular}
  \caption{The corresponding histogram and cumulative histogram of the estimation error when $M = 5$, $10$, $20$ and $30$, respectively, using the two different schemes.}\label{fig:cdf-compare}
\end{figure*}

Figures \ref{fig:cdf-compare}(a) and (b) show the corresponding histograms and cumulative histograms of the estimation errors for $M = 5$, $10$, $20$ and $30$, using the our proposed scheme and the independent noise scheme. The cumulative histograms of our scheme approaches $1$ much slower than those of the independent noise scheme. This indicates that the adversaries obtain less accurate estimations from copies generated by our scheme than from those generated by the independent noise scheme. We also observe that as $M$ increases, the cumulative histograms of our scheme are almost identical as expected; while those by the independent noise scheme approaches the vertical axis, implying estimation errors decrease as adversaries obtain more independently perturbed copies.

In summary, the privacy goal in Section~\ref{ssec:goal} is achieved
in this most severe attacking scenario.

\begin{figure*}[thb]
  \centering
  \begin{tabular}{cc}
  \includegraphics[width=0.89\columnwidth]{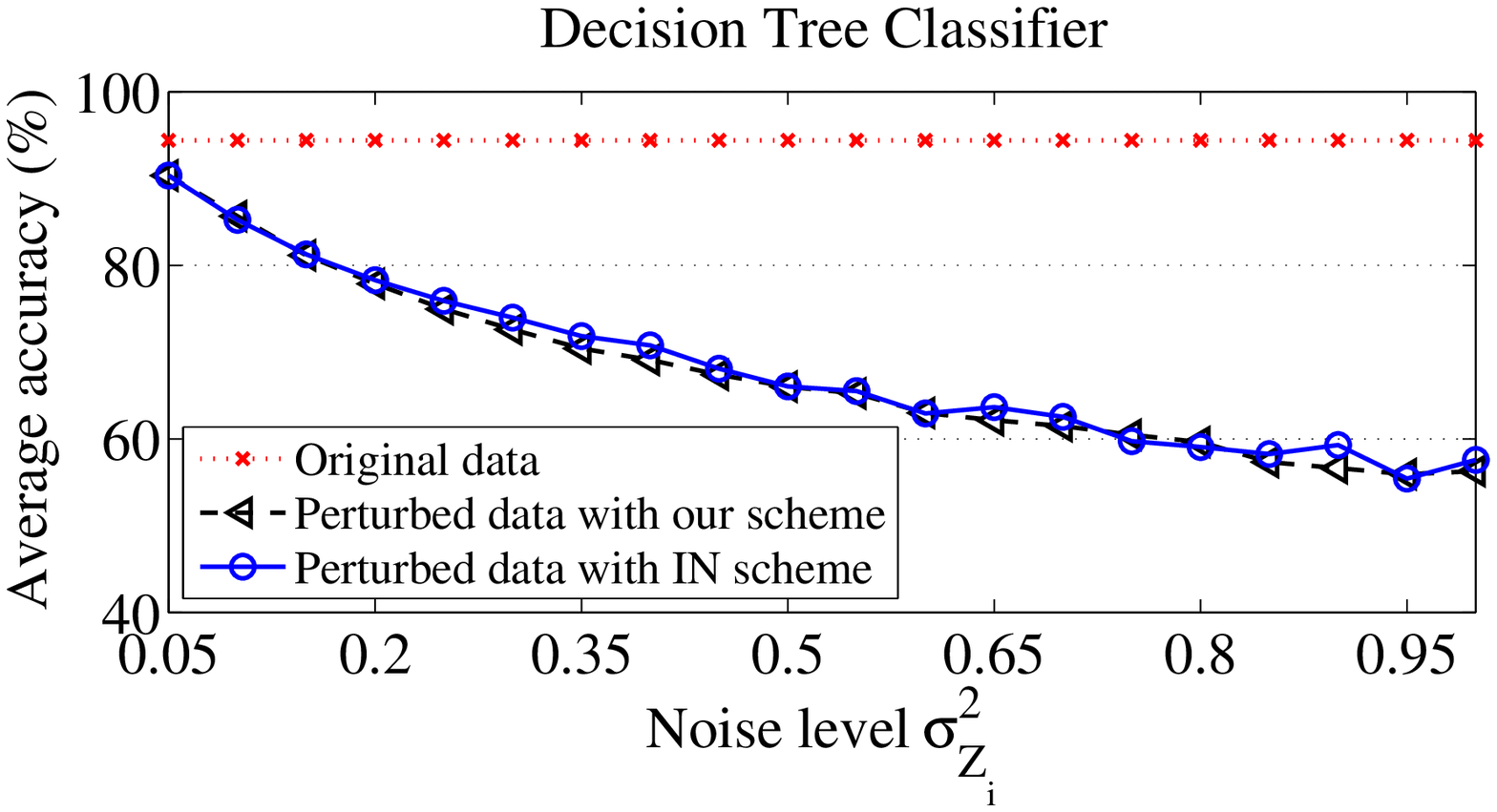} &
  \includegraphics[width=0.89\columnwidth]{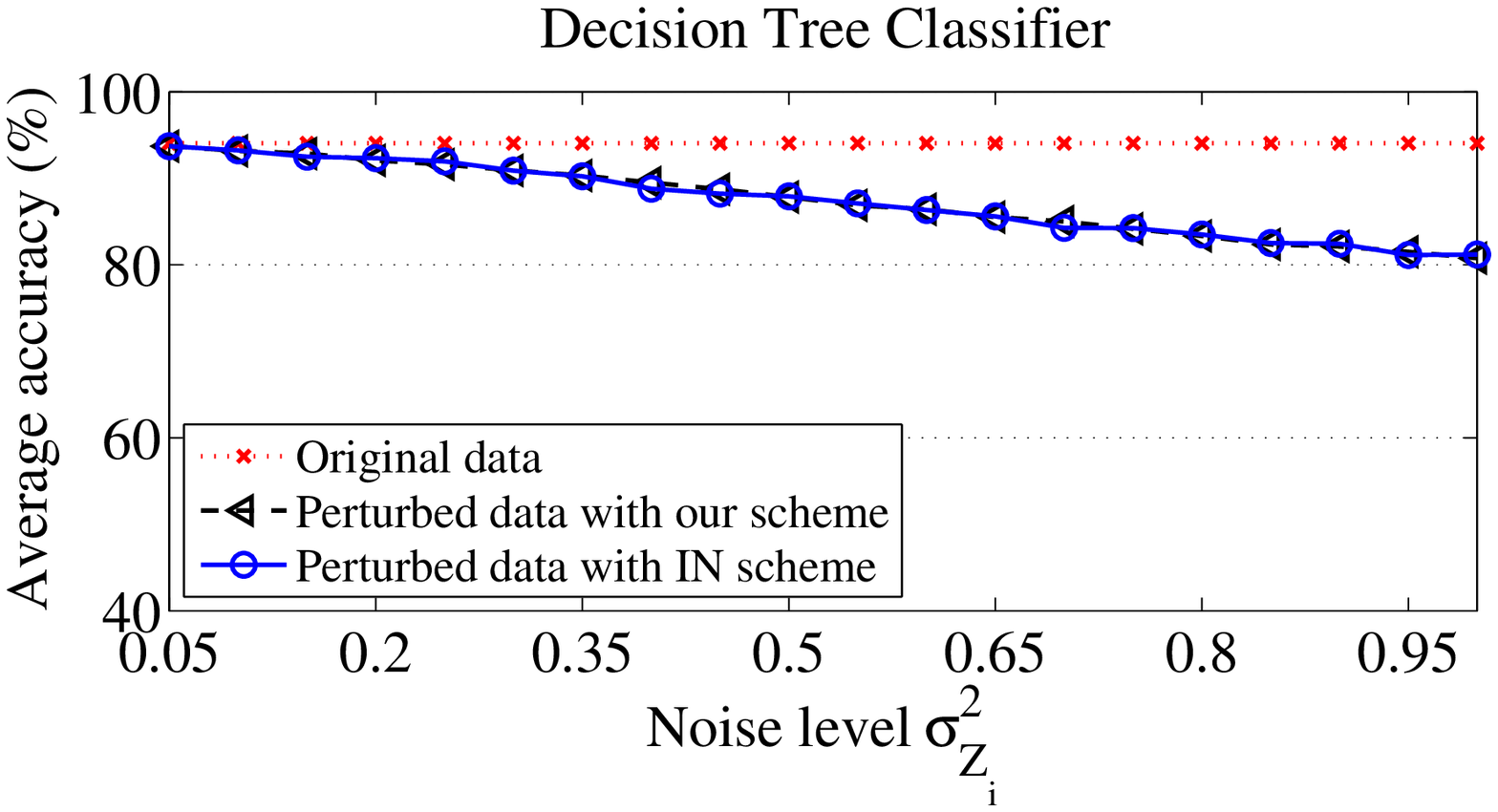} \\
  \includegraphics[width=0.89\columnwidth]{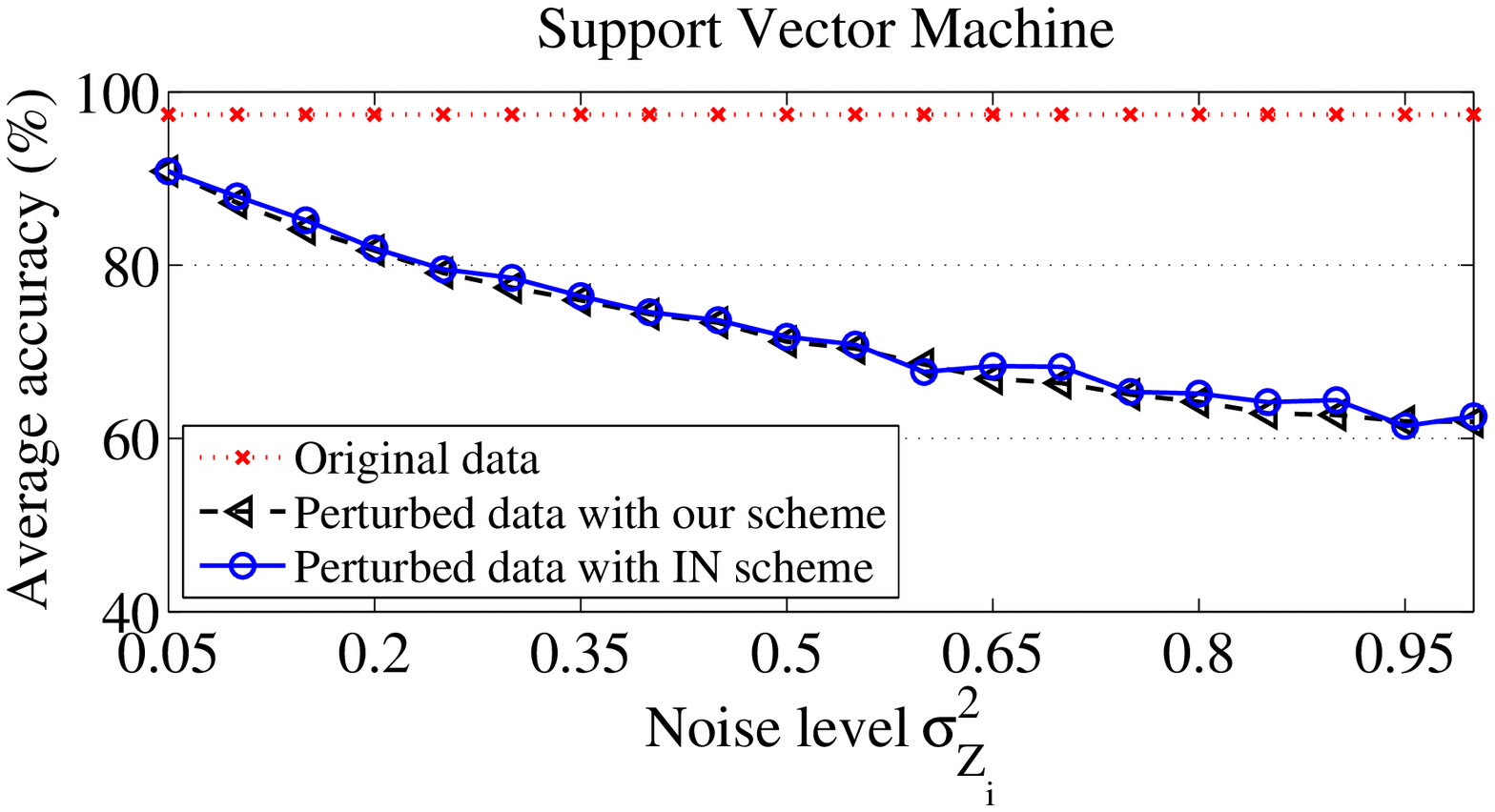} &
  \includegraphics[width=0.89\columnwidth]{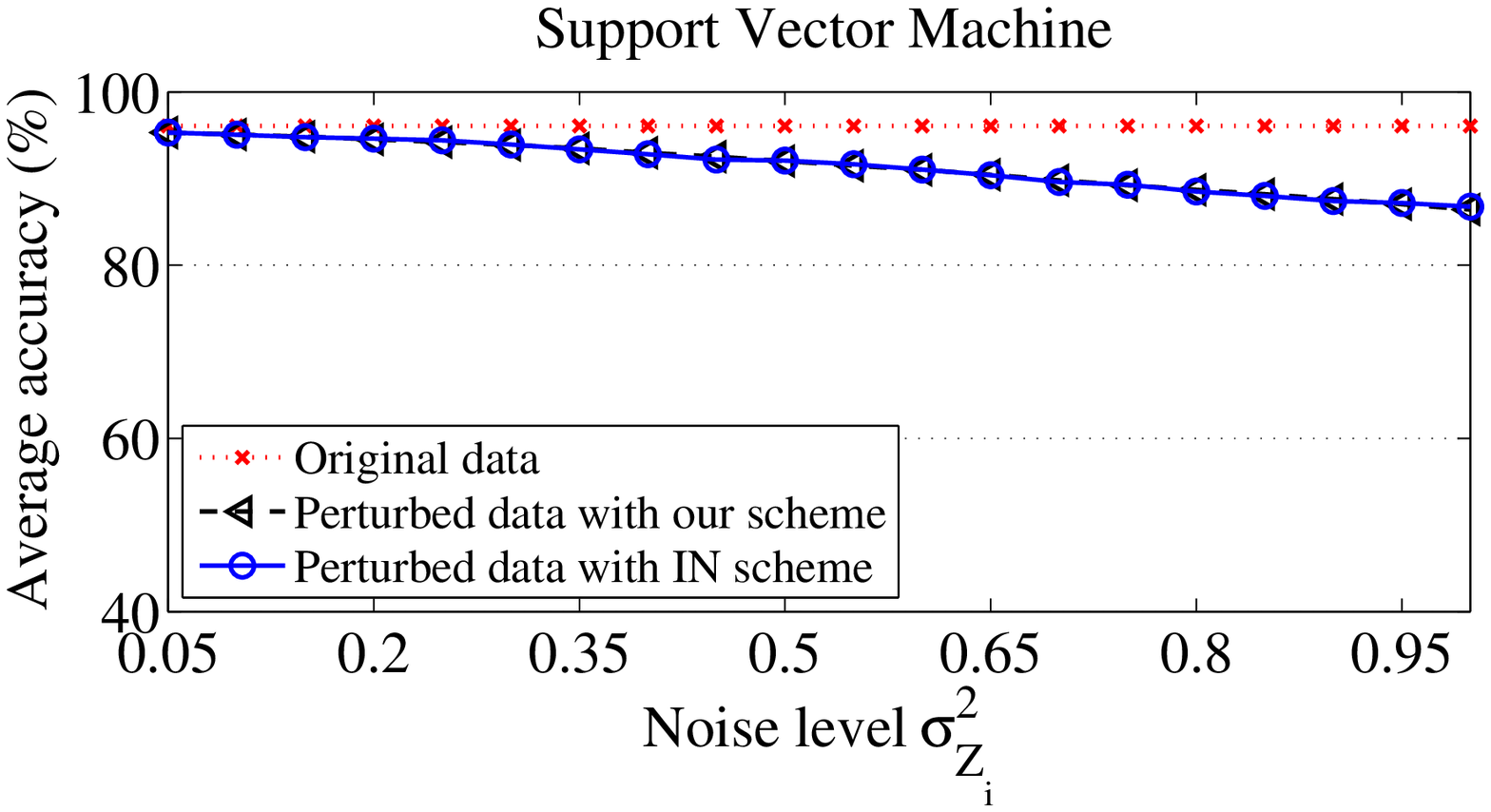} \\
  {(a) Iris Plant database} & {(b) Wisconsin Diagnostic Breast Cancer Database}
  \end{tabular}
  \caption{Comparison of utilities of perturbed copies by different noise addition techniques.
  We show the classification accuracy on the perturbed data at $M = 20$ different noise levels. The Iris Plant database has $150$ tuples with four numerical attributes, and contains three classes of $50$ tuples each. The Wisconsin Diagnostic Breast Cancer database has $699$ tuples with nine numerical attributes, and contains two classes. }\label{fig:utility}
\end{figure*}

We further verify that the perturbed copy by our scheme has the same utility as that by the independent noise scheme, if their trust levels are the same. We use the Iris Plant and Wisconsin Diagnostic Breast Cancer databases from the UCI Machine Learning Repository for the experiment. We measure the utilities with a decision tree classifier and a SVM classifier with radial basis kernel. The average accuracies over $10$-fold cross validation are reported in Fig. \ref{fig:utility}.
As seen from Fig.~\ref{fig:utility}, at all noise levels, the accuracies by the same classifier on the data perturbed by adding independent noise and by properly adding correlated noise following our scheme are identical. Therefore, the perturbed copies at the same trust level by different noise addition techniques have the same utilities.

\subsection{Experiment 2: Scalability Test}
The scalability test is conducted in MATLAB v7.6 on a PC with 2.5GHz CPU and 2GB memory. The attribute \emph{Income} is used as the original data.
We only test Algorithm \ref{alg:A3} as it offers the maximum flexibility in generating perturbed copies and it has the highest time complexity among
our three proposed algorithms.
We use the independent noise scheme with the same settings as a baseline algorithm. Note that this scheme, although with less runtime, is not resistent to diversity attacks.

Theorem \ref{theorem:complexity:alg3} states that to generate one tuple, the time complexity is $O(M^3 + N^3)$. To generate $T$ tuples together, some of the computation can be shared, e.g., generating the covariance matrix of $\zz''$. As a result, the total time complexity to generate $T$ perturbed tuples is $O(M^3 + N^3 + T(M^2 N + M N^2))$, and the average time complexity for one tuple is $O(M^2 N + M N^2)$ for large $T$.

\begin{figure}[tb]
  \centering
    \includegraphics[width=0.91\columnwidth]{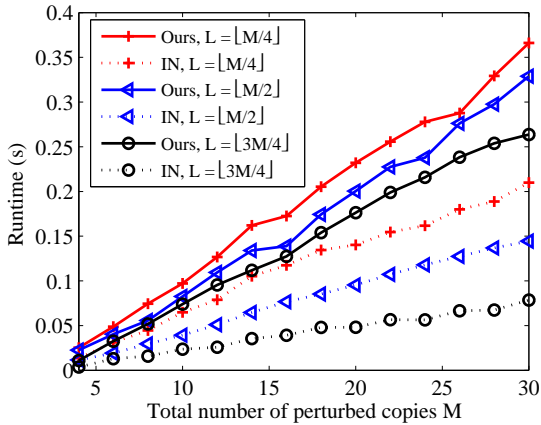}
  \caption{The runtime as a function of the total number of perturbed copies $M$, when the data owner generates $M-L$ perturbed copies each of $10^5$ tuples. The runtime is averaged on $100$ repeated tests.}\label{fig:time1}
\end{figure}

Figure \ref{fig:time1} shows the runtime of Algorithm \ref{alg:A3} as a function of the total number of perturbed copies $M$. For each value of $M$, the data owner generates $M-L$ perturbed copies each of $10^5$ tuples. We set $L= \lfloor M/4 \rfloor$, $\lfloor M/2 \rfloor$, and $\lfloor 3M/4 \rfloor$ respectively. Our observations are three-folded. First, our algorithm is fast. For example, generating $23$ perturbed copies ($M = 30$, $L = \lfloor M/4 \rfloor = 7$) only takes $0.37$ seconds. Second, the actual runtime of Algorithm \ref{alg:A3} we observe only increases approximately linearly in $M$. This observed complexity is much smaller than the theoretical upper bound $O(M^3 + N^3 + M^2 N + M N^2)$ we estimated in Section~\ref{ssec:Space-Time-Complex}. Third, the runtime difference between Algorithm \ref{alg:A3} and the independent noise scheme is considerably small. The time complexity of Algorithm \ref{alg:A3} is the same as that of generating jointly Gaussian noise given the mean and covariance. One of the reasons why the independent noise scheme is marginally faster is that it uses an all-zero mean vector and diagonal covariance matrix.



\section{Conclusion and future work} \label{sec:conclusion}
In this work, we expand the scope of additive perturbation
based PPDM to multi-level trust (MLT), by relaxing an implicit
assumption of single-level trust in exiting work. MLT-PPDM allows
data owners to generate differently perturbed copies of its data for
different trust levels.

The key challenge lies in preventing the data miners from combining copies at different trust levels to jointly reconstruct the original data more accurate than what is allowed by the data owner.

We address this challenge by properly correlating noise across copies at
different trust levels. We prove that if we design the noise covariance
matrix to have corner-wave property, then
data miners will have no diversity gain in their joint reconstruction of
the original data. We verify our claim and demonstrate the
effectiveness of our solution through numerical evaluation.

Last but not the least, our solution allows data owners to
generate perturbed copies of its data at arbitrary trust levels
on-demand. This property offers the data owner maximum flexibility.

We believe that multi-level trust privacy preserving data mining can find many applications. Our work takes the initial step to
enable MLT-PPDM services.

Many interesting and important directions are worth
exploring. For example, it is not clear how to expand the scope of other approaches in the area of partial information hiding, such as
random rotation based data perturbation, $k$-anonymity, and retention
replacement, to multi-level trust. It is also of great interest 
to extend our approach to handle evolving data streams.

As with most existing work on perturbation based PPDM, our work is limited in the sense that it considers only linear attacks. More powerful adversaries may apply nonlinear techniques to derive original data and recover more information. Studying the MLT-PPDM problem under this adversarial model is an interesting future direction.


\section*{Acknowledgement} The authors would like to thank Dr. Xiaokui Xiao for discussions related to the time complexity analysis.

\bibliographystyle{IEEEtran}
\bibliography{IEEEabrv,mining}
\newpage
\begin{biography}{Yaping Li}
received her B.S. degree from the Department of Computer Science at State University of New York at Stony Brook and her Ph.D. degree from the Department of Electrical Engineering and Computer Sciences at University of California at Berkeley. She is currently a postdoc researcher at the Department of Information Engineering of the Chinese University of Hong Kong. Her research interests include database privacy, secure network coding, and applications for secure coprocessors.
\end{biography}
 
\begin{biography}{Minghua Chen}
received his B.Eng. and M.S. degrees from the Department of Electronics Engineering at Tsinghua University in 1999 and 2001, respectively. He received his Ph.D. degree from the Department of Electrical Engineering and Computer Sciences at University of California at Berkeley in 2006. He spent one year visiting Microsoft Research Redmond as a Postdoc Researcher. He joined the Department of Information Engineering, the Chinese University of Hong Kong, in 2007, where he currently is an Assistant Professor. He wrote the book ``General Framework for Flow Control in Wireless Networks'' with Avideh Zakhor in 2008, and the book ``IPv6 Principle and Practice'' with Haisang Wu, Maoke Chen, Xinwei Hu, and Cheng Yan in 2000. He received the Eli Jury award from UC Berkeley in 2007 (presented to a graduate student or recent alumnus for outstanding achievement in the area of Systems, Communications, Control, or Signal Processing), the ICME Best Paper Award in 2009, and the IEEE Transactions on Multimedia Prize Paper Award in 2009. His research interests include complex systems and networked systems, distributed and stochastic network optimization and control, multimedia networking, p2p networking, wireless networking, multi-level trust data privacy, network coding and secure network communications.
\end{biography}

\begin{biography}{Qiwei Li}

received the B.Eng. degree in electronic engineering from Tsinghua University in 2008, and the M.Phil. degree in information engineering from The Chinese University of Hong Kong in 2010. Currently, he is a Research Assistant working at the Department of Statistics, The Chinese University of Hong Kong. His research interests focus on Bioinformatics, including repeats detection, motif discovery, microarray data analysis, gene network inference, etc.

\end{biography}

\begin{biography}{Wei Zhang}
received the B.Eng. degree in electronic engineering from
Tsinghua University, Beijing, in 2007, and the M.Phil. degree in
information engineering from the Chinese University of Hong Kong in
2009. He is currently a Ph.D. candidate in the Department of
Information Engineering of the Chinese University of Hong Kong. His
research interests include machine learning and its applications to computer vision, image processing, and data mining.
\end{biography} 

~

\newpage
\appendices

\section{Deduction of Equation \ref{eqn:llse1}} \label{appendix:proof.eqn.llse1}
Assume that the LLSE estimate $\hat{\db}(\noisyDB) = A \noisyDB + b$, where $A$
and $b$ are parameters.
LLSE minimizes the square errors between the estimated data $\hat{\db}(\noisyDB)$
and the original data $\db$, i.e.,
$$
\begin{array}{lcl}
J & = & {1\over2} E\{Tr[(\db - \hat{\db}(\noisyDB))(\db - \hat{\db}(\noisyDB))^T]\} \\
 & = & {1\over2} E\{Tr[(\db - A \noisyDB - b)(\db - A \noisyDB - b)^T]\}.
\end{array}
$$
As $J$ is a quadratic function of $A$ and $b$, the optimal values of $A$ and $b$ satisfy that
$$
\begin{array}{lcl}
{\partial J \over \partial A} & = & -
E[(\db - \hat{\db}(\noisyDB))\noisyDB^T] = 0, \\
{\partial J \over \partial b} & = & -
E[\db - \hat{\db}(\noisyDB)] = 0.
\end{array}
$$
The above equations are called the orthogonality principle, from which
$$\begin{array}{lcl}
A & = & K_{XY}K_{Y}^{-1}, \\
b & = & E[X] - K_{XY}K_{Y}^{-1} E[Y].
\end{array}$$
Thus, we have
$$\hat{\db}(\noisyDB) = K_{XY}K_{Y}^{-1} [Y - E[Y]] + E[X],$$
where $E[X] = E[Y] = \mu_X$.

\section{Proof of Theorem~\ref{theorem:case.independence}
} \label{appendix:proof.theorem:case.independence}
We first prove the \emph{if} part of the theorem. From the
covariance matrix of $\NoiseS$ and $\NoiseM$, we know that $E[\NoiseS\NoiseM] = \varianceGS \sigma_X^2$.
Therefore,
\begin{equation}\label{eqn:case.indepent}
    E[\NoiseS(\NoiseM-\NoiseS)]
    = E[\NoiseS\NoiseM] - E[\NoiseS^2]
    = \varianceGS \sigma_X^2 - \varianceGS \sigma_X^2
    = 0,
\end{equation}
suggesting that $\NoiseS$ and $\NoiseM-\NoiseS$ are
linearly independent.


Meanwhile, by definition of jointly Gaussian, $\NoiseM-\NoiseS$ is also a Gaussian
random variable. For Gaussian variables $\NoiseS$ and $\NoiseM-\NoiseS$, linear independence implies independence.

We now prove the \emph{only if} part of the theorem.
We observe that $\NoiseM = \NoiseS + (\NoiseM-\NoiseS)$ is sum of two independent
Gaussian random variables. Thus, $\NoiseM$ and $\NoiseS$ are jointly
Gaussian by definition, and we also have $    E[\NoiseM\NoiseS] = E[\NoiseS\NoiseM] = \varianceGS \sigma_X^2$.
It follows that their covariance matrix is as follows:
\[
\left[\begin{array}{cc} \varianceGS \sigma_X^2 & \varianceGS \sigma_X^2\\ \varianceGS \sigma_X^2 &
\varianceGM \sigma_X^2
\end{array} \right].
\]

\section{Proof of Theorem~\ref{theorem:case.suff.cond}
} \label{appendix:proof.theorem:case.suff.cond}
By Theorem~\ref{theorem:independence}, if $\NoiseS$ and $\NoiseM$ satisfy that $\NoiseS$ and $\NoiseM-\NoiseS$ are independent, then their covariance matrix, denoted by $K_C$, must be given by
\[
K_C = \left[\begin{array}{cc} \varianceGS \sigma_X^2 & \varianceGS \sigma_X^2\\ \varianceGS \sigma_X^2 &
\varianceGM \sigma_X^2
\end{array} \right].
\]

Based on $Y_1$, the LLSE estimation of $\db$ has an estimation error of
\begin{equation}\label{eqn:apdx.b.1}
    \sigma_X^2 - \frac{\sigma_X^2}{1 +\sigma_1^2} = \frac{\sigma^2_X}{1+1/\sigma^2_1
        },
\end{equation}
which can be computed using Equation~\ref{eqn:LLSE.Y.est.err}.

Similarly, based on both $Y_1$ and $Y_2$, the LLSE estimation of $\db$ has an estimation error of
\[
\left[\frac{1}{\sigma_X^2} + \left[\begin{array}{cc}
1 & 1\end{array}\right] K_C^{-1}\left[\begin{array}{c}
1\\
1\end{array}\right] \right]^{-1}.
\]
After simplification, the above estimation error is exactly the one shown in Equation~\ref{eqn:apdx.b.1}. Thus, Equation~\ref{eqn:case.goal} holds.

\section{Proof of Theorem~\ref{theorem:ind.noise}
} \label{appendix:proof.theorem:ind.noise}
If $Z_i, 1\leq i\leq \levels$ are independent to each other, then $K_{\zz}$ is given by
\[
K_{\zz} = \left[\begin{array}{cccc}
\sigma_{Z_{1}}^{2}K_{X} & 0 & \cdots & 0\\
0 & \sigma_{Z_{2}}^{2}K_{X} & \cdots & 0\\
\vdots & \vdots & \ddots & \vdots\\
0 & 0 & \cdots & \sigma_{Z_{M}}^{2}K_{X}\end{array}\right].
\]
By Equation~\ref{eqn:LLSE.yy.est.err}, the estimation errors are the diagonal terms of the following matrix
\[
\left[K_X^{-1} + H^T K^{-1}_{\zz} H\right]^{-1} = \left(1+\sum_{i=1}^{M}\frac{1}{\sigma^2_{Z_i}} \right)^{-1} K_X.
\]

\section{Proof of Theorem~\ref{theorem:suff.cond}
} \label{appendix:proof.theorem:suff.cond}
By the definition of distortion and the result shown in Equation~\ref{eqn:LLSE.yy.est.err}, we have
\begin{eqnarray*}
    \dis(\db, \hat{\db}(\yy)) 
   &=&  \frac{1}{N}Tr\left(\left[K_X^{-1} + H^T K^{-1}_{\zz} H\right]^{-1}\right),
\end{eqnarray*}
and for $i=1,\ldots, \levels$,
\begin{eqnarray*}
    \dis(\db, \hat{\db}(Y_i)) 
   &=&  \frac{\sigma^2_{Z_i}}{1+\sigma^2_{Z_i}} \frac{1}{N}Tr\left(K_X\right).
\end{eqnarray*}

Two observations can be made for the above two equations. First, we must have $\dis(\db, \hat{\db}(Y_i))<\dis(\db, \hat{\db}(Y_{i+1}))$ due to the assumption on $\sigma_{Z_i}$ in Equation~\ref{eqn:asmp:sigma_z}, and
\[
    \min_{i=1,\ldots, \levels}\dis(\db, \hat{\db}(Y_i)) = \dis(\db, \hat{\db}(Y_1)) = \frac{\sigma_{Z_1}^2}{\sigma_{Z_1}^2+1}\frac{Tr(K_X)}{N}.
\]
Second, the proof is complete if we can show that
\begin{equation}\label{eqn:apdx.c.1}
    H^TK^{-1}_{\zz}H =  K^{-1}_{Z_1}.
\end{equation}
This obviously holds for the case of $M=1$.

Rewrite $K_{\zz}$ as the following form
\[
    K_{\zz} = \left[\begin{array}{cccc}
K_{Z_{1}} & K_{Z_{1}} & \cdots & K_{Z_{1}}\\
K_{Z_{1}} & \frac{\sigma^2_{Z_2}}{\sigma^2_{Z_1}}K_{Z_{1}} & \cdots & \frac{\sigma^2_{Z_2}}{\sigma^2_{Z_1}}K_{Z_{1}}\\
\vdots & \vdots & \ddots & \vdots\\
K_{Z_{1}} & \frac{\sigma^2_{Z_2}}{\sigma^2_{Z_1}}K_{Z_{1}} & \cdots & \frac{\sigma^2_{Z_M}}{\sigma^2_{Z_1}}K_{Z_{1}}\end{array}\right].
\]
We find its inverse following a standard process. We perform row
operation to the matrix $[K_{\zz}\;|\; I]$ until it has the form
$[I\;|\; A]$. Then matrix $A$ is $K^{-1}_{\zz}$. Note the structure
of $K_{\zz}$ makes this process pretty straightforward and easy.

Following above process, we find the expression of $K^{-1}_{\zz}$ for the case of $M\geq 2$ as follows:
\begin{equation}  \label{eqn:app:invKz}
\small
\left[\begin{array}{ccccc}
\frac{c_{1}\sigma_{Z_{2}}^{2}}{\sigma_{Z_{1}}^{2}}K_{Z_{1}}^{-1} & -c_{1}K_{Z_{1}}^{-1} & 0 & \cdots & 0\\
-c_{1}K_{Z_{1}}^{-1} & (c_{1}+c_{2})K_{Z_{1}}^{-1} & -c_{2}K_{Z_{1}}^{-1} & \cdots & 0\\
0 & -c_{2}K_{Z_{1}}^{-1} & (c_{2}+c_{3})K_{Z_{1}}^{-1} & \cdots & 0\\
\vdots & \vdots & \vdots & \ddots & \vdots\\
0 & 0 & 0 & \cdots & c_{M-1}K_{Z_{1}}^{-1}\end{array}\right],
\end{equation}
where
\[
c_i = \frac{1}{\sigma^2_{Z_{i+1}}/\sigma^2_{Z_1}-\sigma^2_{Z_{i}}/\sigma^2_{Z_1}}, \;\;\;\; 1\leq i\leq M-1.
\]
It is straightforward to verify the product of $K_{\zz}$ and the above matrix is an identity matrix.

Noticing that $K_{\zz}^{-1}$ only have non-zero entries in the main diagonal and two adjacent diagonals, and that its column and row sums are zero except the first row and column, we have
\[
    H^TK^{-1}_{\zz}H = \left[\begin{array}{cccc}
K_{Z_{1}}^{-1} & 0 & \cdots & 0\end{array}\right]\left[\begin{array}{c}
I_{N}\\
\vdots\\
I_{N}\end{array}\right] = K_{Z_{1}}^{-1},
\]
and the proof is complete.

\section{Proof of Theorem~\ref{theorem:independence}
} \label{appendix:proof.theorem:independence}
We first prove the if part of the theorem. Since $Z_1$ to $Z_M$ are jointly Gaussian variables, $Z_1$, and $(Z_{i}-Z_{i-1})$ for  are also jointly Gaussian variables. This is because any linear combination of them is simply another linear combination of $Z_1$ to $Z_M$, and is thus a Gaussian. For jointly Gaussian variables, they are mutually independent if their covariance matrix is a diagonal matrix. This can be easily verified by evaluating their joint distribution.

From the covariance matrix of $\zz$, we know that for $j>i$, $E[\Noise_{i}\Noise_{j}^T] = K_{Z_i}$. For $2\leq i< j\leq M$, we have
\begin{eqnarray*}
    && E[(\Noise_i-\Noise_{i-1})(\Noise_j-\Noise_{j-1})^T] \\
    &=& E[\Noise_i\Noise_j^T] - E[Z_i Z_{j-1}^T] -E[Z_{i-1}Z_j^T] + E[Z_{i-1}Z_{j-1}^T]\\
    &=& K_{Z_i} - K_{Z_i} -K_{Z_{i-1}} + K_{Z_{i-1}} = 0.
\end{eqnarray*}
We also have for $2\leq i\leq M$,
\begin{eqnarray*}
    E[Z_1(\Noise_i-\Noise_{i-1})^T]
    &=& E[Z_1\Noise_i^T] - E[Z_1 Z_{i-1}]^T \\
    &=& K_{Z_1} - K_{Z_1}= 0.
\end{eqnarray*}
As such, we must have the covariance matrix of $Z_1$, and $(Z_{i}-Z_{i-1})$ for  to be diagonal, and they are mutually independent.

We now prove the only if part of the theorem. Since $Z_1$, and $(Z_{i}-Z_{i-1})$ for $i$ from $2$ to $M$ are mutually independent Gaussian variables, we must have $Z_1$ to $Z_M$ to be jointly Gaussian. This is because each of them is simply a linear combination of independent Gaussian variables.

We also have for $j>i$,
\begin{eqnarray*}
    E[\Noise_i\Noise_j^T] &=& E\left[\Noise_i\left(Z_i+\sum_{l=i+1}^j(Z_l-Z_{l-1})\right)^T\right]\\
    &=& E[Z_iZ_i^T]+\sum_{l=i+1}^j E[Z_i(Z_l-Z_{l-1})^T]\\
    &=& K_{Z_i}.
\end{eqnarray*}
It follows that $K_{\zz}$ must have the form as in Equation~\ref{eqn:suff.cond}.

\section{Time and Space Complexity
}
For ease of discussion, we summarize
the time complexity of several basic operations as follows:

\begin{itemize}
  \item {\bf Multiplication of two matrices:} the complexity of multiplication of an $P_1\times P_2$ matrix and an $P_2\times P_3$ matrix is $O(P_1 P_2 P_3)$ by direct computation.

  \item {\bf Cholesky decomposition of $P \times P$ matrices:} time complexity is $O(P^3)$~\cite[pp. 245]{golub1996mc}.
\end{itemize}

\subsection{Proof of Lemma \ref{lem:jointGauss}} \label{appendix:complexity:lemma}
To generate jointly Gaussian random vector $\G$, the standard routine
\cite{gen.gaussian.rvs} generates independent zero-mean unit-variance Gaussian vector
$\G$ and then uses a linear transformation
\begin{equation} \label{eqn:app:LTGauss}
\G = \mu_{\G} + L_{\G} \mathbb{N},
\end{equation}
where $K_{\G} = L_{\G} L_{\G}^T$ is the Cholesky decomposition~\cite[pp. 245]{golub1996mc} of $K_{\G}$.

If both $\Sigma_{\G}$ and $K_0$ are positive semi-definite, we can perform the Cholesky decomposition
as $\Sigma_{\G} = L_{\Sigma} L_{\Sigma}^T$ and $K_0 = L_0 L_0^T$, and then $$K_{\G} 
= (L_{\Sigma} L_{\Sigma}^T) \otimes (L_0 L_0^T) = (L_{\Sigma} \otimes L_0) (L_{\Sigma} \otimes  L_0)^T.$$
Thus the Cholesky decomposition of $K_{\G}$ can be expressed s $L_{\G} = L_{\Sigma} \otimes L_0$.

Following that, Equation \ref{eqn:app:LTGauss} can be written as
\begin{equation} \label{eqn:app:LTGaussKron}
{\G} = \mu_{\G} + (L_{\Sigma} \otimes L_0) \G = \mu_{\G} + \myvec(L_0 \mathbb{N}_{Q \times P}
L_{\Sigma}^T),
\end{equation}
where $\mathbb{N}_{Q \times P}$ is a $Q \times P$ matrix satisfying that $\myvec(\mathbb{N}_{Q \times P}) = \mathbb{N}$.

The total time complexity is the sum of the complexity of generating $PQ$ independent zero-mean unit-variance Gaussian
random variables\footnote{$PQ$ independent zero-mean unit-variance Gaussian
random variables can be generated using a standard algorithm, e.g., \cite{knuth81:ch3}.} $O(PQ)$, the Cholesky decomposition $O(P^3 + Q^3)$,
the matrix multiplication $O(Q^2 P + P^2 Q)$, and the vector addition $O(PQ)$,\footnote{Since the time complexity of generate $PQ$
independent zero-mean unit-variance Gaussian random variables and the vector addition are both $O(PQ)$, which can be bounded by
$O(P^2 Q + P Q^2)$, we omit the complexity of them in the proof of Theorems \ref{theorem:complexity:alg1}--\ref{theorem:complexity:alg3}.} i.e., $O(P^3 + Q^3)$. 

To complete the proof, the remaining part of the proof shows that $\Sigma_{\G}$ is positive semi-definite given that $K_{\G}$ and $K_0$
are covariance matrices and $K_{\G} = \Sigma_{\G} \otimes K_0$.

The definition of positive semi-definite matrices suggests that $F^T K_{\G} F \geq 0$ for an arbitrary column vector $F$.
Without loss of generality, we assume that the element $K_{0}(1,1)$ in the top-left corner of $K_{0}$ is positive. Then we let $F_1$ be composed of all the $(i Q + 1)$-th ($i = 0,...,P-1$)
elements of $F$ and let the other elements of $F$ be zero, and thus
$F_1^T (K_0(1,1) \Sigma_{\G}) F_1 = F^T K_{\G} F \geq 0$. It is straightforward that $\Sigma_{\G}$ is positive semi-definite as $F_1^T \Sigma_{\G} F_1 \geq 0$ for any $F_1$.

\subsection{Proof of Theorem \ref{theorem:complexity:alg1}} \label{appendix:complexity:alg1}
With the technique in the proof of Lemma \ref{lem:jointGauss}, generating $MN$-dimensional jointly Gaussian noise vector ${\zz}$
can use Equation \ref{eqn:app:LTGaussKron}.

In Algorithm 1, the covariance matrix of $\zz$ is $K_\zz = \Sigma_\zz \otimes K_X$, where $$\Sigma_{\zz} = \begin{bmatrix}
\sigma_1^2 & \sigma_1^2 & \cdots & \sigma_1^2 \\
\sigma_1^2 & \sigma_2^2 & \cdots & \sigma_2^2 \\
\vdots & \vdots & \ddots & \vdots \\
\sigma_1^2 & \sigma_2^2 & \cdots & \sigma_M^2 \\
\end{bmatrix}.$$
It is easy to verify that for the Cholesky decomposition $\Sigma_{\zz} = L_{\Sigma} L_{\Sigma}^T$,
$$L_{\Sigma} = U \diag(V_{\sigma}),$$
where $U$ is the lower triangular part (including the diagonal) of all-one
$M \times M$ matrix, $V_\sigma = [\sigma_1, \sqrt{\sigma_2^2 -
\sigma_1^2}, ..., \sqrt{\sigma_M^2 - \sigma_{M-1}^2}]$, and
$\diag(V_\sigma)$ is a diagonal matrix with the vector $V_\sigma$ as
its diagonal.

Thus Equation \ref{eqn:app:LTGaussKron} can be written as
\begin{equation}  \label{eqn:app:alg1Gauss}
{\zz} = \mu_{\zz} + \myvec(L_X (\G_{N \times M} \diag(V_{\sigma})) U^T),
\end{equation}

The matrix multiplication in Equation \ref{eqn:app:alg1Gauss} can
be split into three steps, as shown with the brackets,
with a time complexity $O(MN)$, $O(MN)$, and $O(N^2 M)$,
respectively.

As a result, Algorithm 1 has a time complexity of $O(N^3 + M N^2)$.

The space complexity is $O(M + N^2)$, as Algorithm 1 has to store the noise levels $\sigma_1^2, ..., \sigma_M^2$
and the covariance matrix $K_X$.

\subsection{Proof of Theorem \ref{theorem:complexity:alg2}} \label{appendix:complexity:alg2}
According to Lemma \ref{lem:jointGauss}, when generating one perturbed copy of $X$, the Cholesky decomposition of $K_X$ has a time complexity of $O(N^3)$, and the rest part costs $O(N^2)$.
As the Cholesky decomposition of $K_X$ can be reused for different copies, generating $M$ perturbed copies of $X$ only has a time complexity $O(N^3 + M N^2)$.

Algorithm 2 only requires a memory of size $O(N^2)$ for the covariance matrix $K_X$, as
the noise levels $\sigma_i^2$ ($i = 1, ..., M$) are input sequentially.

\subsection{Proof of Theorem \ref{theorem:complexity:alg3}} \label{appendix:complexity:alg3}
Algorithm 3 first constructs the $(MN)\times(MN)$ matrix $K_Z$
in $O(MN)$ time. It then computes a mean and variance
according to Equations~\ref{eqn:zz''.cond.expt} and~\ref{eqn:zz''.cond.var}.

Note that Equations~\ref{eqn:zz''.cond.expt} and~\ref{eqn:zz''.cond.var} can be written as
$[(\Sigma_{{\zz}'' {\zz}'} \Sigma_{{\zz}'}^{-1}) \otimes I_N] v_1$ and
$(\Sigma_{{\zz}''} - \Sigma_{{\zz}'' {\zz}'} \Sigma_{{\zz}'}^{-1} \Sigma_{{\zz}'' {\zz}'}^T) \otimes K_X$, respectively,
where $K_{\zz''} = \Sigma_{\zz''} \otimes K_X$, $K_{\zz'} = \Sigma_{\zz'} \otimes K_X$ and $K_{\zz''\zz'} = \Sigma_{\zz''\zz'} \otimes K_X$.
$\Sigma_{\zz'}^{-1}$ has been given in an explicit form in Equation \ref{eqn:app:invKz}. So the time complexity of computing the mean and the Kronecker product form of the covariance matrix are $O((M-L)L^2 + (M-L)LN)$ and $O((M-L)L^2 + (M-L)^2 L)$, respectively.

At the end, Algorithm 3 generates $(M - L)$ jointly Gaussian
variables with the computed mean and covariance matrix,
and outputs $(M - L)$ perturbed copies. According to Lemma \ref{lem:jointGauss}, the time complexity
is $O((M-L)^3 + N^3 + (M-L)^2 N + (M-L) N^2)$.

For any value of $L$, the time complexity
of Algorithm 3 is bounded by $O(M^3 + N^3 + M^2 N + M N^2)$,
which can be further simplified to $O(M^3 + N^3)$.

For Algorithm 3, it requires $O(M^2 + N^2)$ memory to store
the covariance matrix $K_{\zz}$.

\end{document}